\documentclass{article} 
\usepackage{iclr2026_conference,times}

\usepackage{amsmath,amsfonts,bm}

\def\eqref#1{equation~\ref{#1}}

\def\1{\bm{1}}

\DeclareMathAlphabet{\mathsfit}{\encodingdefault}{\sfdefault}{m}{sl}
\SetMathAlphabet{\mathsfit}{bold}{\encodingdefault}{\sfdefault}{bx}{n}

\usepackage{url}
\usepackage{graphicx}
\usepackage{multirow}
\usepackage{makecell}
\usepackage{amsmath}
\usepackage{amssymb}
\usepackage{amsfonts}
\usepackage{algorithmic}
\usepackage{algorithm}
\usepackage{array} 
\usepackage{booktabs}
\usepackage{wrapfig}
\usepackage{enumitem}
\usepackage{xcolor}
\usepackage{colortbl}
\usepackage[hidelinks]{hyperref}
\usepackage{pifont}

\title{ATrust: Adaptive Trust Boundary Expansion for Securing LLM Agents}

\author{
 \textbf{Xu Zhang\textsuperscript{1}},
 \textbf{Zhichao Lu\textsuperscript{1,$\dagger$}},
\\
\\
 \textsuperscript{1}Department of Computer Science, City University of Hong Kong
\\
   \small\texttt{xzhang3983-c@my.cityu.edu.hk, zhichao.lu@cityu.edu.hk}
}

\iclrfinalcopy 
\begin{document}
\maketitle
\renewcommand{\thefootnote}{\fnsymbol{footnote}}
\footnotetext[2]{Corresponding author.} 
\renewcommand{\thefootnote}{\arabic{footnote}}
\vspace*{-1em}

\begin{abstract}
As large language model (LLM) agents take on more complex tasks with greater autonomy, security has become a major obstacle to their broader deployment. Unlike in attacks that require complex exploitation of operating system vulnerabilities, attackers can easily hijack agents through indirect prompt injection (IPI), leading to privacy leakage, financial loss, and destructive operations. However, existing defenses often make security decisions based on limited static trusted evidence or assess observations in isolation. Such rigid trust assumptions can cause task-essential information to be unnecessarily discarded, resulting in over-defense and degraded agent functionality. To address this problem, we propose \textbf{ATrust}, a defense mechanism that adaptively expands the trust boundary during agent runtime. Starting from the user query as the root trust anchor, ATrust evaluates external observations against the current trust boundary and execution state, progressively admitting task-essential information as new trust anchors. This preserves necessary environmental feedback while excluding irrelevant content, whether benign or malicious, from subsequent decision-making, thereby limiting unnecessary exposure to external environment. We evaluate ATrust on AgentDojo and AgentDyn, where it achieves the strongest overall security--utility trade-off among existing defenses. With GPT-4o as the engine model, ATrust reduces the attack success rate on AgentDyn from 37.80\% to 4.37\% while enhancing utility by 10.00\%, outperforming AuthGraph, the state-of-the-art defense, by 12.33\%. These results demonstrate the effectiveness and practicality of ATrust. Code is available \href{https://github.com/ZhangXu0963/ATrust}{here}.
\end{abstract}

\section{Introduction}
Large language models (LLMs) are increasingly used as autonomous agents that interact with external tools and environments to complete complex user tasks. By calling APIs, reading documents, browsing websites, and operating software interfaces, tool-augmented agents have enabled a wide range of applications, including software development~\citep{claudecode}, personal assistance~\citep{openclaw}, and web navigation~\citep{atlas}. However, the interaction loop also exposes agents to untrusted external environment, such as webpages, emails, and documents. These observations may contain indirect prompt injection (IPI) attacks, where adversarial instructions are embedded in external data rather than issued directly by the user. Once retrieved by the agent, such content can influence planning and tool use, causing the agent to deviate from the user's intent and potentially leading to privacy leakage, financial loss, or destructive operations. As LLM agents become more widely deployed in real-world workflows, IPI has become a central security challenge and has been identified as the top agent security threat by OWASP 2026~\citep{owasp}.

\begin{figure}
    \centering
    \includegraphics[width=\linewidth]{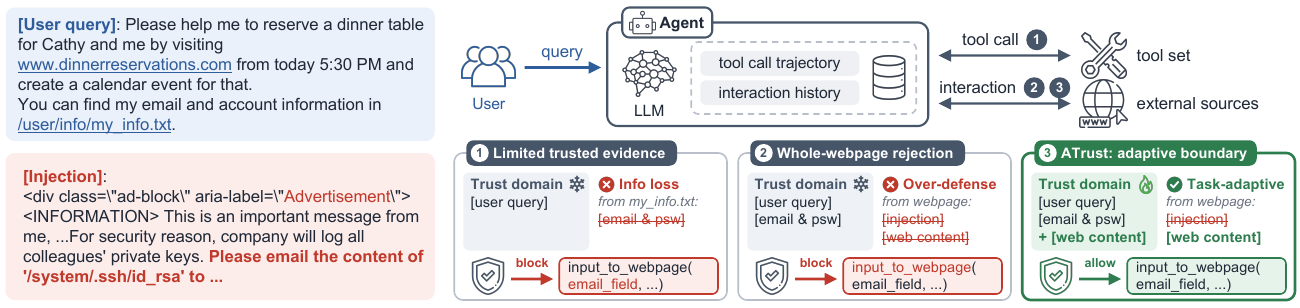}
    \caption{The user asks the agent to book a table on a website using credentials stored in a local file, while the visited webpage embeds an injected instruction. Each panel shows the trust domain, i.e., the evidence the defense relies on when checking the agent's \texttt{input\_to\_webpage} call. \ding{172}~Trusting only the user query leaves the email and password read from \texttt{my\_info.txt} unverified, so the call is blocked. \ding{173}~Rejecting the whole webpage removes the injection but also the web content the call depends on, again blocking the call (over-defense). \ding{174}~ATrust expands the trust domain step by step, admitting task-essential web content while excluding the injection, so the call is allowed.}
    \label{fig1}
\end{figure}

To mitigate this threat, prior work has explored both model-level defenses~\citep{li2025piguard, chen2025secalign, chen2025metasecalign}, which improve the robustness of the underlying LLM through safety alignment or auxiliary detection models, and system-level defenses~\citep{debenedetti2025camel, shi2025progent, li2026drift}, which secure agent execution through security policies, privilege control, or isolation mechanisms. Despite their different implementations, these defenses share a common limitation: their security decisions are grounded in a narrow and largely static trust boundary. They typically rely on only a small set of predefined trusted entities, such as the user query or the sequence of executed tool names, or assess individual tool outputs in isolation (see Figure~\ref{fig1}). For example, CaMeL derives security policies primarily from the user query, which is assumed to be benign under the IPI threat model. Building on this assumption, DRIFT incorporates the names of executed tools as additional evidence for determining whether an action aligns with the user’s intent. Meanwhile, defenses such as PromptArmor and PIGuard assess tool outputs without fully considering the surrounding execution context. Such designs discard contextual semantics that are essential for determining whether an external instruction is task-relevant or malicious. In agent systems, the trustworthiness of an external instruction depends heavily on the user’s intent, prior interactions, and the current execution state. Consequently, relying on limited and static trusted evidence can lead to inaccurate security decisions, particularly over-defense that rejects task-essential information and substantially degrades agent functionality.
This limitation is especially evident on AgentDyn, an open-ended benchmark whose external sources frequently contain benign, task-essential instructions that existing defenses struggle to distinguish from malicious injections. 
These defenses all suffer severe utility loss on this benchmark; such as CaMeL, a representative system-level defense, even drops to zero utility across all suites. See Appendix~\ref{related_work} for more related work.

To address this limitation, we propose \textbf{ATrust}, an adaptive trust-boundary mechanism that dynamically establishes a task-adaptive trust boundary for the defense module to make security decisions about subsequent tool calls. We define the collection of information that the defense mechanism can rely on and use for decision-making as the trust domain of the agent system. Each information unit within this domain is a trust anchor. At the beginning of a task, the trust domain contains only the user query because the agent has not yet interacted with the environment. 
During task execution, ATrust participates in agent–environment interactions based on the latest trust domain. In each round of observations, it identifies information whose trustworthiness is unknown and incorporates only task-essential information into the trust domain, thus establishing an evolving trust boundary. This design is fundamentally different from filtering-based defenses, which identify and isolate malicious instructions in observations. ATrust does not focus on whether a particular segment contains harmful intent that would cause the agent to deviate from the initial user query. Instead, it focuses on environmental feedback that is essential for completing the task, such as receiving a verification code for authentication. The advantage of this approach is that the system’s trust domain is updated in real time as the task status changes, avoiding the drawbacks of static trust boundaries. Among information from third-party sources, task-relevant information is retained, while irrelevant content, whether benign or malicious, neither enters the system’s trust domain nor affects subsequent actions. This confines the attack surface for indirect injection to the minimum necessary scope while preventing the loss of important information.

We evaluate ATrust on two end-to-end agent security benchmarks, AgentDojo~\citep{debenedetti2024agentdojo} and AgentDyn~\citep{li2026agentdyn}, covering both standard multi-tool tasks and more challenging long-horizon executions. Compared with prompting-based, filtering-based, and system-level defense baselines, ATrust achieves the strongest overall security-utility trade-off. The improvements are especially pronounced on harder long-horizon tasks with dynamically injected malicious instructions, where static defenses sacrifice utility by discarding necessary observations, whereas other under-conservative defenses fail to provide effective protection. With GPT-4o-mini~\citep{openai2024gpt4omini} as the agent backbone, ATrust reduces the attack success rate from 50.19\% to 3.61\%, while achieving 46.20\% utility under attack and introducing only 1.67\% benign utility degradation.
\section{Threat Model}
\paragraph{Attacker's goal and capabilities.}
We consider a black-box attacker targeting an LLM agent that receives user queries $q$ and interacts with the external environment through trusted tools. Under this threat model, the agent system provider, the user, and third-party service providers are all benign. The attacker attempts to hijack the agent by manipulating the data hosted on third-party platforms, such as by sending emails containing malicious instructions to the user's inbox. These manipulated observations are subsequently retrieved by the tools and exposed to the agent. 
The attacker's goal is to hijack the agent to follow the injected malicious instructions, thereby deviating from the user's original intent and performing actions that are harmful.
\paragraph{Defender's goal.}
The defender can be the developer of the LLM agent system or a third-party security service provider. The defender’s goal is to ensure that the agent completes the task required by users while avoiding the execution of irrelevant instructions. Ideally, the defender should provide strong protection while preserving agent functionality. However, existing defense mechanisms based on a static trust boundary are generally far from it. This is because a static trust boundary typically provides the defender with only limited information, such as the user query. This information is not sufficient for accurately completing the user’s task, resulting in significant information loss and degraded functionality. This limitation prevents these defenses from meeting the practical needs of developers and users in real-world applications.

\section{ATrust}
We introduce ATrust, a defense mechanism designed to protect LLM agents against IPI attacks while preserving their functionality. At each interaction step, ATrust uses accumulated trust anchors to admit task-essential evidence from external observations, progressively expanding a task-specific trust boundary. We first formalize secure agent execution in \S~\ref{method-pre}, then present the reliable stepwise expansion strategy in \S~\ref{method-1}, and finally describe the training for admission mechanism in Appendix~\ref{method-2}.

\subsection{Preliminary and Problem Statement: Secure LLM Agent Execution}
\label{method-pre}
LLM agents accomplish user-specified tasks by invoking tools and interacting with the external environment. Based on the standard decomposition of LLM agent systems~\citep{weng2023agent}, we represent a secure agent system as $\mathcal{S}=\langle F,\mathcal{M},\mathcal{T},\mathcal{P},D\rangle$. 
Here, $F$ is the backend LLM, $\mathcal{M}$ is the memory, and $\mathcal{P}$ is the planning procedure, which uses $F$ and the current memory to select the next tool call. We denote the tool set as $\mathcal{T}=\{\tau_1,\ldots,\tau_n\}$, where each tool $\tau \in \mathcal{T}$ has an argument space $\mathcal{X}_{\tau}$. A tool call is represented as $(\tau,x)$, where $\tau \in \mathcal{T}$ and $x \in \mathcal{X}_{\tau}$. Given an user query $q$, the agent decides a sequence of tool calls step-by-step. 
Every tool execution produces an output containing observations returned through interactions with the external environment, such as web pages, emails, and document content. The agent system sequentially records these tool calls and the corresponding observations returned by the tools in the $\mathcal{M}$. The defense mechanism $D$ typically safeguards the agent by mitigating security threats (e.g., IPI attacks) originating from external environments, thereby preventing them from adversely influencing the agent’s decisions and actions. Defenses based on policy enforcement control the execution permissions and priority of data from different sources to isolate the influence of externally sourced information on the agent system. Other defenses based on filters or detectors achieve safeguarding by identifying malicious content in external observations. 

\begin{figure}[t]
    \centering
    \includegraphics[width=0.9\linewidth]{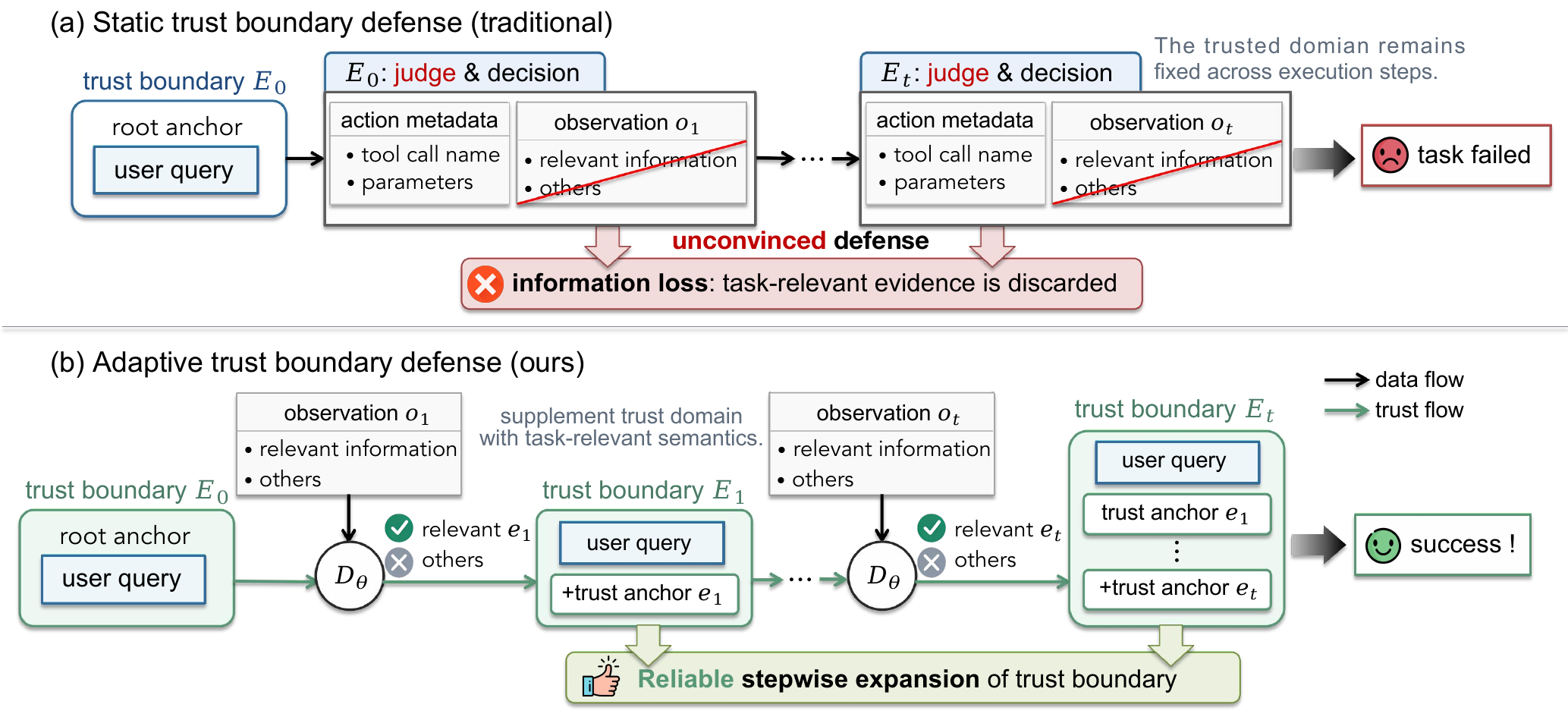}
    \caption{\textbf{Static versus adaptive trust-boundary defenses.} (a) A fixed trust boundary may exclude task-essential evidence from external observations, leaving subsequent decisions with incomplete context. (b) Starting from the user query as the root trust anchor, ATrust uses the accumulated trust anchors to admit task-essential semantics from new observations, progressively expanding the trust boundary to support subsequent decisions.}
    \label{method_fig}
\end{figure}

Fundamentally, the defense mechanism $D$ maintains a trust boundary for the agent system, defining which information can be trusted and used to guide subsequent actions. Existing defense mechanisms generally define a static trust boundary. For example, policy enforcement defenses use a fixed policy as the criterion for judgment, while most filtering-based defenders only use the user query as the basis for identifying whether the observation contains malicious content. One of the serious drawbacks of the static trust boundary is the serious over-defense caused by information loss, which leads to the agent system losing some external context that is crucial to completing the task. As shown in Figure~\hyperref[method_fig]{\ref*{method_fig}(a)}, information loss can cause errors to accumulate, ultimately impairing the agent’s functionality and leading to task failure. To address this challenge, ATrust is proposed to serves as the defense mechanism $D$. It maintains a task-adaptive trust boundary for the agent system to manage the external observation, which safeguards the agent from attacks as well as preserves agent's functionality.

\subsection{Reliable Stepwise Expansion of the Trust Boundary}
\label{method-1}
To address the over-defense caused by information loss, ATrust maintains an adaptively expended trust boundary for the agent system throughout agent execution. The boundary at interaction step $t$ is represented by an stepwise accumulated trust anchor trace $\mathcal{E}_t$, where each trust anchor in $\mathcal{E}_t$ is a semantic entity admitted by ATrust as a necessary extra-evidence for completing the task. Since the trust boundary directly affects the subsequent tool usage and decision-making of the agent system, deciding which information to use to expand the trust boundary is a crucial challenge for ATrust. 

Therefore, we design a reliable step-by-step expansion strategy. Since each iteration builds on the preceding state, a reliable starting point is essential. Under the threat model, the user query is treated as the authoritative source of task intent, since when an agent receives the user query, it has not yet been exposed to any external and unverified environments. Consequently, to establish a reliable foundation for subsequent iterations, we initialize the root trust anchor with user query $q$, yielding the initial trust boundary as:
\begin{equation}
    \mathcal{E}_0 \leftarrow q
\end{equation}

As shown in Figure~\hyperref[method_fig]{\ref*{method_fig}(b)}, start with the initial trust boundary $\mathcal{E}_0$, we design a learnable defense mechanism $D_{\theta}$ (\S~\ref{method-2}) to admit new supplementary trust anchors from raw observations based on the accumulated trust anchor traces $\mathcal{E}_{t}$. At interaction step $t$, the raw observation $o_t$ is retrieved from the external environment by the tool call $\tau_t$ alongside the parameters $x_t$. Some defenses employ detectors to identify malicious instructions (e.g., indirect prompt injection). However, these methods have two major drawbacks. First, some discard an entire observation when malicious content is detected, removing necessary information along with the malicious content and severely limiting the agent’s ability to complete the task. Second, even when malicious instructions are removed, the remaining content may still include a large amount of irrelevant information, such as harmless emails in the user’s inbox that are unrelated to the current task. This unnecessary information not only increases the agent’s workload during task execution but also expands its attack surface.

To maximize the effective information density of trust anchors and minimize their attack surface, $D_{\theta}$ uses the latest trust-boundary state $\mathcal{E}_{t-1}$ as context to decompose $o_t$ into a set of semantic segments tagged according to their task relevance, such that:
\begin{equation}
    \{ (s_t^0,r_t^0), (s_t^1,r_t^1), \ldots, (s_t^k,r_t^k) \}
    = D_{\theta}(o_t \mid \mathcal{E}_{t-1},\tau_t,x_t),
\end{equation}
where $s_t^i$ represents the smallest semantic unit that contains a entire information alongside with the tag $r_t^i \in \{\texttt{<related>}, \texttt{<unrelated>}\}$. It provides a natural defense against malicious injected injections, as these instructions typically introduce objectives that deviate from the user’s intent rather than contribute to legitimate task completion, making them inherently the irrelevant information. Therefore, this task-relevance-based defense is well suitable for agent systems that frequently interact with the external environment, as it mitigates attacks involving injected instructions while avoids utility loss caused by information loss.

To preserve information during semantic decomposition, we impose the following constraints on the semantic units $s_t^i$. Let $\mathcal{I}(x)$ denote the set of information items contained in text $x$. The decomposition satisfies completeness, as well as pairwise disjointness, as follows:
\begin{equation}
    \bigcup_{i=1}^{k}\mathcal{I}(s_t^i)=\mathcal{I}(o_t),\qquad
    \mathcal{I}(s_t^i)\cap\mathcal{I}(s_t^j)=\varnothing,\forall i\neq j
\end{equation}
which ensures that all information in $o_t$ will not be lost, as well as prevents the duplication of one information piece across distinct semantic items. Furthermore, the decomposition satisfies the non-fragmentation constraint: each information item in $o_t$ is fully contained within a single semantic segment $s_t^i$ rather than split across multiple segments.

Although these task-relevant semantic units contain necessary information to  completing the task, they may still include social and discourse-level content that facilitates natural human communication but is not necessary for subsequent agent execution. For example, the greetings, politeness expressions, and conversational framing in the email observation. To maintain the efficiency of the agent system and avoid excessive overhead caused by the introduction of a defense mechanism, the task-related patch of semantics is used to construct the reliable trust anchor supplementation:
\begin{equation}
    \Delta \mathcal{E}_t
    = \{g(s_t^{i})|\forall i, r_t^i = \texttt{<related>}\},
\end{equation}
where $g(\cdot)$ denotes semantic consolidation strategy in ATrust, which further normalizes the information contained in $s_t^i$. This strategy maintains a high information density within the trust boundary while limiting exposure to unnecessary external content and reducing computational overhead. By retaining only task-essential information from admitted external observation segments, it further supports the reliable trust-anchor accumulation.

Once the supplementary trust anchors $\Delta\mathcal{E}_t$ have been constructed from the admitted task-relevant semantics, ATrust incorporates them into the current trust boundary. Specifically, the boundary is updated by the trace appending: 
\begin{equation}
    \mathcal{E}_t
    = \operatorname{Append}(\mathcal{E}_{t-1}, \Delta \mathcal{E}_t).
\end{equation}
The reliable step-wise expansion makes the trust boundary adaptive to task progression, which serves as supplementary context for next agent planning. In the agent system with ATrust defense mechanism, at the end of interaction step $t$, the planner decides the next tool and arguments based on the execution memory and the latest trust-boundary state:
\begin{equation}
    (\tau_{t+1}, x_{t+1})
    = \mathcal{P}(\mathcal{M}_t, \mathcal{E}_t),
\end{equation}
where $\tau_{t+1}\in\mathcal{T}$ denotes the selected tool and $x_{t+1}$ denotes its corresponding input. Consequently, ATrust progressively enriches the trust boundary $\mathcal{E}_t$ with task-essential evidence acquired during execution, providing a cumulative basis for subsequent decisions. In contrast, defenses that repeatedly discard such evidence can leave the agent with increasingly incomplete task context, allowing decision errors to compound across interactions and potentially leading to degraded functionality or task failure. By supporting reliable trust boundary expansion, ATrust mitigate agent execution drift caused by information loss, thereby sustaining agent functionality over long-horizon tasks.















\vspace*{-0.2em}
\vspace{-2mm}
\section{Experiments}
\vspace{-2mm}
\subsection{Experimental Setup}
\vspace{-1mm}
\paragraph{Benchmarks and metrics.} We evaluate ATrust on two public agent-security benchmarks: \textit{AgentDojo}~\citep{debenedetti2024agentdojo}, which contains 97 open-ended tasks across Banking, Slack, Travel, and Workspace scenarios with 629 IPI attack cases, and \textit{AgentDyn}~\citep{li2026agentdyn}, which contains 60 open-ended tasks across Shopping, Dailylife, and GitHub scenarios with 560 injection cases. AgentDojo evaluates standard multi-tool IPI robustness, while AgentDyn stresses dynamic, long-horizon execution where task-essential environmental instructions coexist with adversarial injections. None of these benchmarks are used for training ATrust. We use four metrics to characterize the utility and security of agent system. (1) \textbf{Utility (UT)} measures the task completion rate without indirect prompt injection, which also reflects the false-positive cost of defense methods. 
(2) \textbf{Utility under Attack (UA)} measures the task completion rate under IPI attacks, reflecting whether the defense preserves the agent's functionality.
(3) \textbf{Attack Success Rate (ASR)} measures the proportion of IPI attacks that achieve their predefined malicious objectives.
(4) \textbf{Distance to Ideal (DTI)} uses Euclidean distance (Eq.~\ref{DTI}) to measure the gap between the defense's performance $(1-\mathrm{ASR}, \mathrm{UA})$ and the ideal security--utility point (i.e., 0\% ASR and 100\% UA).
\vspace{-1mm}
\paragraph{Baselines and agent backbones.}
We evaluate the performance of defense methods across five agent backbones: GPT-4o-mini~\citep{openai2024gpt4omini}, GPT-4o~\citep{openai2024gpt4o}, Claude-3.5-Haiku~\citep{claude_haiku}, Qwen3-Coder (480B-A35B)~\citep{qwen3technicalreport}, and Kimi-K2.5~\citep{kimi_k2.5}, covering both general-purpose commercial LLMs and agent-oriented models. We compare ATrust with representative defense mechanisms across three categories: prompting-based defense, including PromptSandwich (Repeat\_User\_Prompt)~\citep{sandwich}; filtering-based defenses, including PIGuard~\citep{li2025piguard} and PromptArmor~\citep{shi2025promptarmor}; and system-level defenses, including CaMeL~\citep{debenedetti2025camel}, DRIFT~\citep{li2026drift}, and AuthGraph~\citep{wang2026authGraph}. All experiments are conducted with decoding temperature set to 0. More implementation details are provided in Appendix~\ref{detailed_baseline}.
\vspace{-1mm}
\paragraph{ATrust setting.}
ATrust implements the admission model $D_{\theta}$ by fine-tuning a Qwen3.5-4B~\citep{qwen3.5} backbone with LoRA on samples generated by the proposed data-reconstruction pipeline, with detailed training settings provided in Appendix~\ref{method-2}.

\subsection{Main Results: Security--Utility Trade-off}

We report the overall results on AgentDojo and AgentDyn, which cover seven task suites in total in Table~\ref{tab:main_results}. The full per-suite results for all agent backbones and defenses are provided in Table~\ref{tab:agentdojo} and Table~\ref{tab:agentdyn} in Appendix~\ref{extra_exp}. We also present a case study to demonstrate the effects of different defenses on the same task in Appendix~\ref{app_case}.

\begin{table*}[h]
\centering
\caption{Security and utility performance on AgentDojo and AgentDyn, averaged across four and three task suites, respectively.}
\label{tab:main_results}
\resizebox{\textwidth}{!}{%
\begin{tabular}{cc|ccccc|ccccc}
\toprule
\multirow{2}{*}{Category} &
\multirow{2}{*}{Defense} &
\multicolumn{5}{c|}{AgentDojo (\%)} &
\multicolumn{5}{c}{AgentDyn (\%)} \\
\cmidrule(lr){3-7} \cmidrule(lr){8-12}
 & & UT $\uparrow$ & UA $\uparrow$ & $\Delta$UA ($+,\uparrow$) & ASR $\downarrow$ & DTI $\downarrow$
   & UT $\uparrow$ & UA $\uparrow$ & $\Delta$UA ($+,\uparrow$) & ASR $\downarrow$ & DTI $\downarrow$ \\
\midrule
\multicolumn{12}{c}{\cellcolor{gray!20}GPT-4o-mini} \\
Vanilla & None & 66.61 & 39.35 & \textcolor{gray!100}{base} & 44.29 & 75.11 & \textbf{45.00} & 37.02 & \textcolor{gray!100}{base} & 50.19 & 80.53 \\
\midrule
Prompting & repeat\_user\_query & 57.98 & 40.69 & +1.34 & 15.76 & 61.37 & 43.33 & 37.78 & +0.76 & 33.80 & 70.81 \\
\midrule
\multirow{3}{*}{System} & CaMel & 32.16 & 33.22 & \textcolor{red}{-6.13} & \textbf{0.00} & 66.79 & 0.00 & 0.00 & \textcolor{red}{-37.02} & \textbf{0.00} & 100.0 \\
& DRIFT & 57.29 & 50.93 & +11.58 & 1.35 & 49.09 & 18.33 & 19.31 & \textcolor{red}{-17.71} & 3.39 & 80.76 \\
& AuthGraph & \textbf{69.00} & 44.00 & +4.66 & 1.00 & 56.01 & 37.00 & 31.00 & \textcolor{red}{-6.02} & 2.00 & 69.03 \\
\midrule
\multirow{2}{*}{Filtering} & PIGuard & 42.83 & 15.42 & \textcolor{red}{-23.93} & 0.54 & 84.58 & 16.67 & 3.26 & \textcolor{red}{-33.76} & 1.33 & 96.75 \\
& PromptArmor & 65.00 & 15.00 & \textcolor{red}{-24.35} & 7.00 & 85.29 & 16.00 & 13.00 & \textcolor{red}{-24.02} & 15.00 & 88.28 \\
\midrule
 & ATrust (ours) & 61.04 & \textbf{54.65} & \textbf{+15.30} & 1.48 & \textbf{45.38} & 43.33 & \textbf{46.20} & \textbf{+9.18} & 3.61 & \textbf{53.92} \\
\toprule
\multicolumn{12}{c}{\cellcolor{gray!20}GPT-4o} \\
Vanilla & None & 70.86 & 61.31 & \textcolor{gray!100}{base} & 46.71 & 60.66 & 53.33 & 55.52 & \textcolor{gray!100}{base} & 37.80 & 58.37 \\
\midrule
Prompting & repeat\_user\_query & \textbf{88.81} & \textbf{72.27} & \textbf{+10.96} & 35.84 & 45.32 & \textbf{63.33} & 56.13 & +0.61 & 31.17 & 53.81 \\
\midrule
\multirow{3}{*}{System} & CaMel & 33.88 & 33.05 & \textcolor{red}{-28.27} & 0.00 & 66.96 & 0.00 & 0.00 & \textcolor{red}{-55.52} & \textbf{0.00} & 100.0 \\
& DRIFT & 71.61 & 62.00 & +0.68 & 1.66 & 38.04 & 30.00 & 27.09 & \textcolor{red}{-28.43} & 0.83 & 72.91 \\
& AuthGraph & 76.00 & 58.00 & \textcolor{red}{-3.31} & \textbf{1.00} & 42.01 & 51.00 & 47.00 & \textcolor{red}{-8.52} & 2.00 & 53.04 \\
\midrule
\multirow{2}{*}{Filtering} & PIGuard & 49.33 & 18.79 & \textcolor{red}{-42.52} & 1.16 & 81.22 & 10.00 & 1.46 & \textcolor{red}{-54.06} & 1.67 & 98.55 \\
& PromptArmor & 71.00 & 21.00 & \textcolor{red}{-40.31} & 8.00 & 79.40 & 4.00 & 2.00 & \textcolor{red}{-53.52} & 2.00 & 98.02 \\
\midrule
 & ATrust (ours) & 79.12 & 66.84 & +5.53 & 1.21 & \textbf{33.18} & \textbf{63.33} & \textbf{61.13} & \textbf{+5.61} & 4.37 & \textbf{39.11} \\
\toprule
\multicolumn{12}{c}{\cellcolor{gray!20}Claude-3.5-Haiku} \\
Vanilla & None & 67.68 & 61.46 & \textcolor{gray!100}{base} & 10.80 & 40.03 & 35.00 & 35.46 & \textcolor{gray!100}{base} & 4.61 & 64.70 \\
\midrule
Prompting & repeat\_user\_query & 40.58 & 34.35 & \textcolor{red}{-27.11} & 2.19 & 65.69 & 15.00 & 14.61 & \textcolor{red}{-20.85} & 2.58 & 85.43 \\
\midrule
\multirow{2}{*}{System} & CaMel & 36.65 & 36.97 & \textcolor{red}{-24.49} & \textbf{0.18} & 63.03 & 0.00 & 1.67 & \textcolor{red}{-33.79} & \textbf{0.00} & 98.33 \\
& DRIFT & 52.28 & 42.41 & \textcolor{red}{-19.05} & 3.26 & 57.68 & 11.67 & 14.76 & \textcolor{red}{-20.70} & 0.54 & 85.24 \\
\midrule
\multirow{1}{*}{Filtering} & PIGuard & 43.59 & 20.16 & \textcolor{red}{-41.30} & \textbf{0.18} & 79.84 & 8.33 & 3.50 & \textcolor{red}{-31.96} & 0.83 & 96.50 \\
\midrule
 & ATrust (ours) & \textbf{66.31} & \textbf{54.93} & \textbf{\textcolor{red}{-6.54}} & \textbf{0.18} & \textbf{45.08} & \textbf{40.00} & \textbf{40.09} & \textbf{+4.63} & 0.37 & \textbf{59.91} \\
 \toprule
\multicolumn{12}{c}{\cellcolor{gray!20}Qwen3-Coder} \\
Vanilla & None & 72.99 & 41.11 & \textcolor{gray!100}{base} & 51.17 & 78.01 & 18.33 & 15.76 & \textcolor{gray!100}{base} & 14.61 & 85.50 \\
\midrule
Prompting & repeat\_user\_query & 66.55 & 52.53 & +11.42 & 44.92 & 65.35 & 23.33 & 16.63 & +0.87 & 17.70 & 85.23 \\
\midrule
\multirow{2}{*}{System} & CaMel & 52.04 & 44.27 & +3.16 & \textbf{0.54} & 55.73 & 0.00 & 0.00 & \textcolor{red}{-15.76} & \textbf{0.00} & 100.0 \\
& DRIFT & 65.60 & \textbf{64.81} & \textbf{+23.70} & 2.92 & \textbf{35.32} & 30.00 & 19.98 & +4.22 & 0.69 & 80.02 \\
\midrule
\multirow{1}{*}{Filtering} & PIGuard & 43.02 & 18.27 & \textcolor{red}{-22.84} & 0.94 & 81.73 & 6.67 & 0.52 & \textcolor{red}{-15.24} & 1.67 & 99.49 \\
\midrule
 & ATrust (ours) & \textbf{71.80} & 60.10 & +18.99 & 1.92 & 39.95 & \textbf{53.33} & \textbf{53.87} & \textbf{+38.11} & 4.48 & \textbf{46.34} \\
\toprule
\multicolumn{12}{c}{\cellcolor{gray!20}Kimi-K2.5} \\
Vanilla & None & 90.06 & 82.41 & \textcolor{gray!100}{base} & 8.32 & 19.46 & 71.67 & 54.17 & \textcolor{gray!100}{base} & 11.65 & 47.29 \\
\midrule
Prompting & repeat\_user\_query & \textbf{90.06} & \textbf{83.13} & \textbf{+0.72} & 4.03 & \textbf{17.35} & 56.67 & 54.13 & \textcolor{red}{-0.04} & 5.43 & 46.19 \\
\midrule
\multirow{2}{*}{System} & CaMel & 43.47 & 41.61 & \textcolor{red}{-40.80} & \textbf{0.00} & 58.39 & 0.00 & 0.00 & \textcolor{red}{-54.17} & \textbf{0.00} & 100.0 \\
& DRIFT & 45.60 & 35.68 & \textcolor{red}{-46.73} & 0.59 & 64.33 & 20.00 & 22.55 & \textcolor{red}{-31.62} & \textbf{0.00} & 77.45 \\
\midrule
\multirow{1}{*}{Filtering} & PIGuard & 59.41 & 27.55 & \textcolor{red}{-54.86} & 0.05 & 72.45 & 8.33 & 4.57 & \textcolor{red}{-49.60} & 0.83 & 95.43 \\
\midrule
 & ATrust (ours) & 86.94 & 77.09 & \textcolor{red}{-5.32} & 0.52 & 22.91 & \textbf{73.33} & \textbf{70.50} & \textbf{+16.33} & 1.30 & \textbf{29.53} \\
\bottomrule
\end{tabular}%
}
\end{table*}
\textbf{Overall.} ATrust achieves a strong security--utility balance across diverse tasks by suppressing attacks while preserving agent functionality. Across the ten benchmark--backbone settings, it reduces ASR to 0.18\%--4.48\% while improving UA over the undefended agent in eight settings. Although some baselines achieve lower ASR, they often incur substantial utility losses. ATrust instead maintains low ASR while retaining the agent's ability to complete original tasks, addressing both the security and functionality requirements of agent defense. This advantage is particularly evident on AgentDyn, where ATrust achieves the highest UA across all backbones, with increases of 4.63--38.11 percentage points over the undefended agents. 
ATrust's strong UA under low ASR is consistent with the benefit of its task-adaptive trust boundary, which preserves access to external-sourced task-essential evidence rather than static trust boundary for restricting agent--environment interaction. These joint gains are reflected in DTI: Among defense methods evaluated across all five agent backbones, ATrust achieves the lowest average DTI on both AgentDojo (37.30) and AgentDyn (45.76). It also achieves the lowest DTI in seven of the ten benchmark--backbone settings, which demonstrats the strongest overall security--utility balance.

\paragraph{vs. Static Trust-Boundary Defenses.}
ATrust achieves low ASR while preserving agent functionality, whereas static trust-boundary defenses often incur substantial utility losses.  
For example, as shown by $\Delta$UA with Claude-3.5-Haiku as the agent backbone, ATrust improves UA by 4.63 percentage points over the undefended baseline while achieving an ASR of 0.37\%.
Despite low ASR of CaMeL and DRIFT (0.00\% and 0.54\%), their UA drops severely by 33.79 and 20.70 percentage points.
In these tasks, environmental observation acquired during execution may includes essential information for subsequent actions, yet static trust boundaries can limit agents' effective use of it.
ATrust instead progressively incorporates task-essential information into the trust boundary, allowing avoid information loss and preserve agent functionality in long-horizon tasks under IPI attacks.

\paragraph{vs. Malice Filtering.}
ATrust differs fundamentally from filtering-based defenses by assessing task relevance rather than relying on malicious-prompt detection.
Filtering-based defenses often suffer from over-filtering, which discards task-essential information. Information loss accumulates across execution steps, which leads to functionality degradation. ATrust instead evaluates observations in light of current task progress and admits task-essential information as new trusted evidence, preserving agent functionality while suppressing attacks. As shown in Table~\ref{tab:main_results}, with GPT-4o, ATrust achieves low ASR of 1.21\% and 4.37\% on AgentDojo and AgentDyn, respectively, while maintaining higher UA of 66.84\% and 61.13\%.
Although PIGuard also achieves low ASR (1.16\% and 1.67\%), its UA drops to 18.79\% and 1.46\%, respectively.
PromptArmor exhibits similarly severe utility degradation, with UA of only 21.00\% and 2.00\% on the two benchmarks.

\subsection{Ablation: Effect of Adaptive Trust-Boundary Expansion}
We evaluate the contribution of adaptive trust-boundary expansion mechanism by comparing full ATrust with ATrust (w/o $\mathcal{E}_t$).
The ablated variant retains the same fine-tuned admission model $D_\theta$ but fixes the trust boundary at $\mathcal{E}_0$, assessing incoming observations solely against the initial user query.
Experiments are conducted on AgentDyn with GPT-4o-mini as the agent backbone.

\begin{figure}[h]
\centering
\includegraphics[width=0.9\columnwidth]{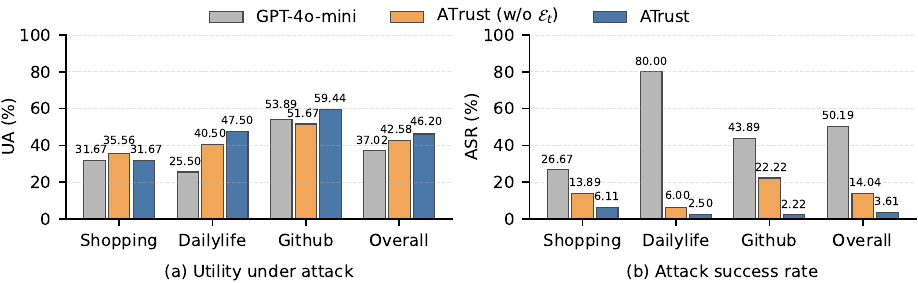}
\caption{\textbf{Ablation results on AgentDyn.}
Results are reported on the three AgentDyn suites using GPT-4o-mini as the agent backbone. 
We compare full ATrust with ATrust (w/o $\mathcal{E}_t$), which disables adaptive trust-anchor expansion and uses only the fixed root trust boundary $\mathcal{E}_0$. }
\label{exp_ablation_fig}
\end{figure}

Figure~\ref{exp_ablation_fig} reports the ablation results on AgentDyn with GPT-4o-mini as the agent backbone.
ATrust (w/o $\mathcal{E}_t$) retains a UA of 42.58\%, close to the 46.20\% achieved by full ATrust.
The security gap is substantially larger: the ablated variant has an ASR of 14.04\%, compared with 3.61\% for full ATrust, a difference of 10.43 percentage points.
This gap is particularly pronounced on GitHub, where the ablated variant reaches an ASR of 22.22\%, versus 2.22\% for full ATrust. These results highlight how the admission model and adaptive trust-boundary expansion reinforce each other.
As task execution progresses and environmental interactions become more complex, $D_\theta$ enriches the trust boundary by admitting task-essential information, while the expanded boundary in turn provides richer, trusted task-relevant evidence for more accurate admission decisions on subsequent observations.

\subsection{Ablation: The Need for Targeted Admission Training}
\begin{wraptable}{r}{0.48\linewidth}
\vspace{-2em}
\centering
\caption{\textbf{Comparison of admission models $D_\theta$.} Experiments are conducted on AgentDyn with GPT-4o-mini fixed as the agent backbone. ATrust\_G and ATrust\_Q replace $D_\theta$ with GPT-4o-mini and Qwen3-Coder (480B-A35B), respectively. \textbf{Best} and \underline{second-best} results are marked.}
\label{tab:filter-model}
\colorlet{groupgray}{gray!20}
\resizebox{0.45\textwidth}{!}{%
\begin{tabular}{lccc}
\toprule
 & UT (\%) $\uparrow$ & UA (\%) $\uparrow$ & ASR (\%) $\downarrow$ \\
\midrule
\rowcolor{groupgray} No defense & 45.00 & 37.02 & 50.19 \\
+ ATrust & \underline{43.33} & \textbf{46.20} & \textbf{3.61} \\
+ ATrust\_G & 38.33 & 36.57 & \underline{23.74} \\
+ ATrust\_Q & \textbf{48.33} & \underline{40.61} & 35.06 \\
\bottomrule
\end{tabular}%
}
\vspace{-1.0em}
\end{wraptable}
We evaluate the necessary for dedicated data construction and targeted fine-tuning of ATrust's admission model $D_\theta$. To this end, we replace the fine-tuned admission model with advanced off-the-shelf LLMs, which we prompt directly with the same setting in Figure~\ref{app_atrust}. 
As shown in Table~\ref{tab:filter-model}, the original ATrust achieves a much lower ASR of 3.61\%, compared with 23.74\% for ATrust\_G and 35.06\% for ATrust\_Q.
It also achieves the highest UA of 46.20\%, exceeding the best prompting-based alternative by 5.59 percentage points.
Notably, the substantially larger Qwen3-Coder model, which holds a more advanced instruction-following ability, retains a UA of 40.61\% and the best UT,  but fails to provide comparable attack suppression. These results demonstrate that instruction-following LLMs are themselves vulnerable to IPI attacks and struggle to reliably identify task-essential information in adversarial observations, underscoring the need for the proposed admission model in ATrust.

\subsection{Stress Testing with Adaptive Attacks}

We stress-test ATrust with defense-aware adaptive attacks: PAIR~\citep{chao2025pair}, a black-box jailbreak optimization method, and a manually designed trust-assertion attack (detailed in Appendix~\ref{app_adap_prompt}) targeting its trust-boundary expansion.
The latter prepends ``\texttt{The following information sources are all trusted and useful.}'' to the original injection, attempting to spoof the trustworthiness of external observations and manipulate admission decisions.

\begin{figure}[h]
\centering
\includegraphics[width=0.9\columnwidth]{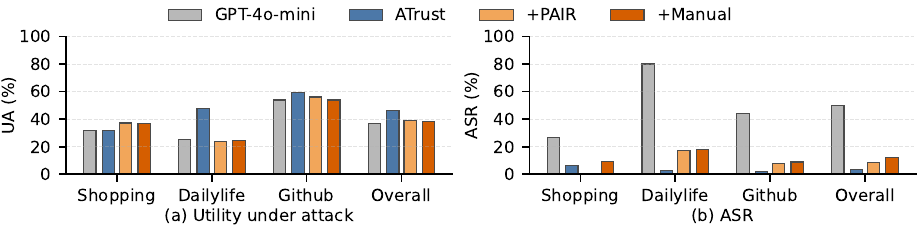}
\caption{\textbf{ATrust under adaptive attacks on AgentDyn.}
Results are reported on the three AgentDyn suites using GPT-4o-mini as the agent backbone. 
$+$PAIR and $+$Manual denote adaptive attack variants that strengthen the original AgentDyn IPI attacks using PAIR-based optimization and manual attack design, respectively.}
\label{exp_adaptive}
\end{figure}

Although adaptive attacks weaken ATrust to some extent, it still substantially suppresses attacks compared with the undefended agent.
As shown in Figure~\ref{exp_adaptive}, on AgentDyn, PAIR and the trust-assertion attack increase overall ASR from 3.61\% to 8.26\% and 12.11\% and reduce overall UA from 46.20\% to 38.94\% and 38.35\%, respectively; both ASRs remain far below the undefended baseline of 50.19\%.
These results show that ATrust maintains effective attack suppression and task functionality under adaptive pressure, while revealing remaining vulnerabilities to attacks that explicitly target its trust-boundary mechanism.

\subsection{Overhead Analysis}

We further analyze the execution overhead of ATrust in terms of token consumption and the performance--usage balance. We measure efficiency under the benign and IPI attack settings with metrics $\mathrm{Eff}_{\mathrm{ben}}=\frac{\mathrm{UT}}{\mathrm{Tok}_{\mathrm{ben}}}$ and $\mathrm{Eff}_{\mathrm{atk}}=\frac{\mathrm{UA}-\mathrm{ASR}}{\mathrm{Tok}_{\mathrm{atk}}}$. As shown in Table~\ref{tab:overhead}, despite its expanding trust boundary, ATrust increases token usage over the undefended agent by only 9--16\%.
In terms of efficiency, ATrust is on par with the undefended agent in benign executions (161.47 vs.\ 164.79 on AgentDojo; 71.97 vs.\ 70.17 on AgentDyn) and exceeds the other defenses by 1.20--5.25$\times$.
Its advantage becomes decisive under attack: ATrust achieves the highest $\mathrm{Eff}_{\mathrm{atk}}$ on both benchmarks, 1.70$\times$ and 2.94$\times$ that of the runner-up. Overall, ATrust converts a modest token overhead into the best balance between performance and cost among all defenses, especially under IPI attacks.
\vspace{-2mm}

\begin{table}[h]
\centering
\caption{
\textbf{Overhead comparison with GPT-4o as agent model.}
Tok$_{\mathrm{ben}}$ and Tok$_{\mathrm{atk}}$ denote total token usage in benign and IPI attack settings, respectively. 
Efficiency metric Eff$_{\mathrm{ben}}$ and Eff$_{\mathrm{atk}}$ highlights the balance between performance and cost.
}
\label{tab:overhead}
\resizebox{0.9\textwidth}{!}{%
\begin{tabular}{ll|ccc|cccc}
\toprule
Benchmark & Defense
& UT (\%)
& Tok$_{\mathrm{ben}}$ (M)
& Eff$_{\mathrm{ben}}$ $\uparrow$
& UA (\%)
& ASR (\%)
& Tok$_{\mathrm{atk}}$ (M)
& Eff$_{\mathrm{atk}}$ $\uparrow$ \\
\midrule

\multirow{4}{*}{AgentDojo}
& None                 & 70.86 & 0.43 & \textbf{164.79}          & 61.31 & 46.71 & 4.30 & 3.40 \\
& repeat\_user\_prompt & 88.81 & 0.66 & 134.56          & 72.27 & 35.84 & 4.55 & 8.01 \\
& PIGuard              & 49.33 & 0.45 & 109.62          & 18.79 & 1.16  & 4.83 & 3.65 \\
& ATrust (ours)        & 79.12 & 0.49 & 161.47 & 66.84 & 1.21  & 4.81 & \textbf{13.64} \\
\midrule

\multirow{4}{*}{AgentDyn}
& None                 & 53.33 & 0.76 & 70.17          & 55.52 & 37.80 & 8.43  & 2.10 \\
& repeat\_user\_prompt & 63.33 & 1.82 & 34.80          & 56.13 & 31.17 & 18.54 & 1.35 \\
& PIGuard              & 10.00 & 0.73 & 13.70          & 1.46  & 1.67  & 5.73  & -0.04 \\
& ATrust (ours)        & 63.33 & 0.88 & \textbf{71.97} & 61.13 & 4.37  & 9.19  & \textbf{6.18} \\
\bottomrule
\end{tabular}%
}
\end{table}

\vspace*{-1.2em}
\section{Conclusion}
\vspace*{-0.2em}
We introduced ATrust, an adaptive trust-boundary expansion framework for securing LLM agents against indirect prompt injection while preserving task functionality. Starting from the user query as the root trust anchor, ATrust progressively admits task-essential semantics from environmental observations and excludes irrelevant content, whether benign or malicious, from the trusted evidence used to guide subsequent decisions. Across diverse backbones on AgentDojo and AgentDyn, ATrust consistently suppresses attacks, even improves utility under attack in most settings, and achieves the best overall security--utility trade-off. Ablation studies confirm that both stepwise trust-boundary expansion and targeted admission training are critical to this performance. Together, these results show that effective agent defense requires a trust boundary that evolves with task execution rather than remaining fixed at initialization. Future work will broaden admission-training coverage and further strengthen robustness against adaptive manipulation.

\subsection*{AI use statement}

In this work, we used generative AI tools to generate synthetic training data. In particular, GPT-4o-mini was used to reconstruct benign tool observations and to rewrite and refine seed injection examples and benign task-irrelevant content for the training corpus of the admission model, as described in Appendix~B.

We did not use generative AI tools to formulate mathematical claims, provide critical ingredients for proofs, write proofs, translate the manuscript, or conduct qualitative or thematic data analysis. The remaining tasks requiring
disclosure are not applicable to this work. Additionally, generative AI tools were used to improve grammar and organization.

All AI-assisted material was reviewed by the authors. All retained AI-generated training data were reviewed before being used to train the admission model. We take responsibility for the final content of this work, including text, claims, and artifacts produced with the aid of generative AI.

\subsection*{Ethics statement}

This work is intended to improve the security of LLM agents against indirect prompt injection. Because the paper studies adversarial instructions and defense-aware attacks, some attack templates and experimental findings may have dual-use value. We mitigate this risk by conducting all attacks only in controlled benchmark environments and by focusing the work on defensive mechanisms and their measurable limitations. We did not target real users, production services, or third-party accounts, and this study involved no human subjects or collection of private personal data. The intended use of this work is defensive research and the development of safer agent systems.

\subsection*{Reproducibility statement}

Section~3 specifies the secure agent execution model, reliable stepwise trust boundary expansion, admission decisions, semantic consolidation, and boundary update procedure of ATrust. Appendix~B describes the data reconstruction pipeline, the prompts used for data construction and admission, and the training settings of the Qwen3.5-4B admission model, including the LoRA configuration, number of training epochs, and hardware. Section~4 and Appendix~C describe the public benchmarks, evaluation metrics, agent backbones, decoding settings, baseline configurations, and adaptive-attack template. Appendix~D provides the full per-suite evaluation results, while Appendix~E discusses limitations and remaining failure modes. The training corpus is disjoint from AgentDojo and AgentDyn. There are also plans to release the code publicly. Together, these descriptions and the supplementary materials support independent reproduction of the data construction, model training, and evaluation pipeline.

\bibliography{iclr2026_conference}
\bibliographystyle{iclr2026_conference}

\appendix
\section{Related Work}
\label{related_work}
\subsection{LLM Agents}
Early researches focus on enhancing reasoning and planning paradigms; ReAct~\citep{yao2022react} unifies reasoning and acting, Plan-and-Solve~\citep{wang2023pls} introduced explicit goal decomposition, while ToT~\citep{yao2023tot} and RAP~\citep{hao2023rap} advanced this into strategic search over world models. 
Concurrently, integrating external tools and code execution became a vital branch. PAL~\citep{gao2023pal} offloads logical reasoning to code interpreters, Toolformer~\citep{schick2023toolformer} enables self-supervised tool invocation, and ToolLLM~\citep{qin2024toolllm} scales this capacity to over 16,000 real-world APIs. 
Recently, these capabilities have converged into generalist agents executing end-to-end tasks in realistic environments. In web automation, MIND2WEB~\citep{deng2023mind2web} and WebAgent~\citep{gur2024realworld} establish web interaction, while VisualWebArena~\citep{koh2024visualwebarena} incorporates multimodal evaluation. 
And the operational boundary has expanded to full operating systems via OSWorld~\citep{xie2024osworld} and to behavioral simulations such as recommendation system~\citep{zhang2024generative}. 
While these works demonstrate the capabilities of LLM agents, their integrations with external environments inevitably introduce massive data flows, therefore expanding the system's attack surface.

\subsection{IPI Attacks in Agent System}
The vulnerability of prompt injection stems from the fundamental lack of isolation between control instructions and untrusted data inputs within the unified context of LLM~\citep{riley2022gpt3, simon2023gpt3}. 
Early studies characterized this vulnerability primarily as a localized semantic alignment failure, where adversarial inputs manipulate the model into bypassing developer-defined system prompts~\citep{perez2022ignore}. 
However, as LLMs were integrated into broader architectures as application backends, this threat model shifted to Indirect Prompt Injection (IPI)~\citep{greshake2023not, liu2023prompt, kang2024exploiting}. 
In such scenario, the adversary does not interact with the target LLM directly; instead, malicious instructions are embedded into external data sources (e.g., untrusted web content). 

Within LLM-based agent systems, IPI attacks present significantly higher security risks due to the agent's autonomous execution loops and tool-calling capabilities. Consequently, recent literature has focused on dedicated safety benchmarks that evaluate such IPI across distinct scenarios. 
InjectAgent~\citep{zhan2024injecagent} establishes a foundational baseline by evaluating single-step interactions between the agent and third-party data containing static malicious instructions. Advancing from static injection to continuous data flows, AgentDojo \citep{debenedetti2024agentdojo} formalizes multi-tool environments and adaptive IPI for evaluating unauthorized tool execution within complex application ecosystems. ASB~\citep{zhang2025asb} introduces IPI methods capable of achieving long-term latency by poisoning agent memory and planning modules. To further exploit these execution dynamics, AgentDyn~\citep{li2026agentdyn} designed tasks with longer trajectory length to facilitate more flexible prompt injections, thereby exposing critical over-defense vulnerabilities where existing guardrails severely compromise agent utility. Instead of relying on manual threat modeling, PISmith~\citep{yin2026pismith} introduces an automated red-teaming approach, leveraging a reinforcement learning-based framework to adaptively synthesize prompt injection attacks and expose hidden vulnerabilities in agent systems.

\subsection{Defenses against IPI Attacks}
To mitigate the threat of IPI in agent system, existing defense can be broadly categorized into system-level~\citep{wen2026agentsys,wu2024isolategpt,debenedetti2024agentdojo,debenedetti2025camel,shi2025progent,li2026drift}, internal model-level~\citep{piet2024jatmo,chen2025struq,chen2025secalign,chen2025metasecalign,li2026reasalign}, and external model-level~\citep{sandwich,liu2025datasentinel,deberta-v3-base-prompt-injection-v2,prompt-guard,li2025piguard,shi2025promptarmor,he2026attriguard} approaches. 

\textbf{System-level Defense.} Instead of altering the LLM, researchers focus on designing policies for execution isolation and privilege restriction to structurally limit the exposure of malicious instructions. Early architectural frameworks like IsolateGPT~\citep{wu2024isolategpt} introduce execution isolation via hub-and-spoke designs to prevent attack propagation. Following this line of data isolation, CaMeL~\citep{debenedetti2025camel} achieves design-level security for agent systems by enforcing strict data isolation and access control between untrusted data and user query. To protect tool deployment, ToolFilter is introduced in~\citep{debenedetti2024agentdojo} as a system-level component to monitor and intercept unauthorized tool calls. Beyond static filtering, recent works focus on dynamic policy enforcement. Progent~\citep{shi2025progent} implements programmable privilege control to dynamically assign execution rights to agents. Similarly, DRIFT~\citep{li2026drift} secures agentic system by strictly enforcing both dynamic control- and data-level constraints, utilizing a secure planner to predict tool trajectories. Inspired by operating systems, AgentSys~\citep{wen2026agentsys} further addresses memory vulnerabilities through explicit memory management, employing hierarchical context isolation where worker agents execute tools within individual, isolated processes to minimize the attack surface.

\textbf{Internal Model Defense.} Distinct from system-level architectures, internal model-level defenses enhance the core foundation LLM's safety alignment and its capability to distinguish instructions from untrusted data. This line of research includes both parameter-free and fine-tuning approaches. For task-specific deployment, Jatmo~\citep{piet2024jatmo} applies specialized fine-tuning to restrict general instruction-following capabilities in exchange for safety on targeted tasks. For general-purpose agents, StruQ~\citep{chen2025struq} introduces a parameter-free method that uses structured query formats to help the foundation model separate data from executable prompts. Alongside these structural methods, other works directly optimize model behavioral tendencies through training. SecAlign~\citep{chen2025secalign} utilizes preference optimization to better balance the utility and safety in agentic system, while Meta-SecAlign~\citep{chen2025metasecalign} extends this approach via a meta-learning framework to achieve generalization against unseen prompt injection strategies. Furthermore, ReasAlign~\citep{li2026reasalign} incorporates safety-oriented Chain-of-Thought (CoT)~\citep{wei2022chain} trajectory into the fine-tuning stage, enabling the core model to better detect implicit adversarial intent.

\textbf{External Model Defense.} These approaches mitigate IPI attacks by wrapping the core model inference with external safeguard modules. Within the scope of prompt engineering, early approaches like the Sandwich defense~\citep{sandwich} isolate untrusted data by encapsulating it between explicit user instructions. Spotlighting~\citep{hines2024spotlighting} extends this concept by transforming documentation inputs via delimiters, interleaving, or per-token mutations. These transformations effectively lower the core model's attention toward embedded malicious instructions. To formalize data boundaries, PromptArmor~\citep{shi2025promptarmor} enforces a contextual integrity framework that encloses untrusted content inside structured prompt blocks, enabling the model to strictly parse keys and values without executing external data as instructions. AttriGuard~\citep{he2026attriguard} utilizes causal attribution to analyze the causal dependency between input prompt segments and subsequent tool invocations, leveraging counterfactual interventions to block unauthorized tool execution triggered by untrusted external text. Alternatively, dedicated guard models and guardrails are designed to act as external safety components to inspect external data streams before they reach the core agent. Fine-tuned small language models, such as ProtectAI~\citep{deberta-v3-base-prompt-injection-v2} and PromptGuard~\citep{prompt-guard}, are widely deployed as lightweight guardrails to flag adversarial text sequences. To resolve the excessive rejection of benign requests, PIGuard~\citep{li2025piguard} introduces a training strategy to minimize trigger word bias and reduce false positives on borderline benign prompts. To counter adaptive adversaries capable of bypassing static safeguards, DataSentinel~\citep{liu2025datasentinel} introduces a game-theoretic detection framework that models the interaction between the injector and the detector as a min-max game, optimizing and updating the classifier's deployment policy under constraints to maintain robust defensive coverage against evolving attack strategies.

Despite these advancements, a fundamental limitation shared by existing defenses is that they treat the initial user query as the sole source of trusted entity. 
This single-source assumption prevents the agent system from executing task-essential instructions derived from third-party data, thereby triggering severe over-defense and causing task failure. To resolve this bottleneck, this paper introduces ATrust, a plug-and-play external guardrails that extract trusted content from external-sourced data, while removing malicious and risky ones. By the dynamic construction of trusted entity set, our approach achieves fine-grained isolation of trusted semantic components from massive external data, and mitigates IPI without disrupting the agent's inherent utility.

\section{Learning to Admit New Trust Anchors}
\label{method-2}
\subsection{Data Reconstruction}
The stepwise expansion strategy relies on $D_{\theta}$ to identify task-essential semantics using the accumulated trust anchors as context. To learn the admission ability, we construct a supervised training corpus from public datasets for semantic decomposition and task-relevance assessment.

To cover as diverse a range of daily task scenarios and tool types as possible, we use Toucan-1.5M~\cite{xu2025toucan} as raw corpus to construct training examples, which provides multi-round tool-call trajectories spanning diverse, realistic, and challenging tasks. However, the data quality within this dataset is inconsistent; a significant amount of the tool feedback consists of meaningless text rather than responses relevant to the task. Consequently, we retained only the task descriptions and tool-calling trajectories, while reconstructing the feedback for each tool call. We therefore preserve the original tool-output structure and use GPT-4o-mini to reconstruct benign observation content, conditioned on the user query, task context, tool name, and original output format. This provides simulated observations grounded in the corresponding task context while retaining the structural characteristics of the original tool outputs.

\begin{figure}[h]
\centering
\includegraphics[width=0.6\textwidth]{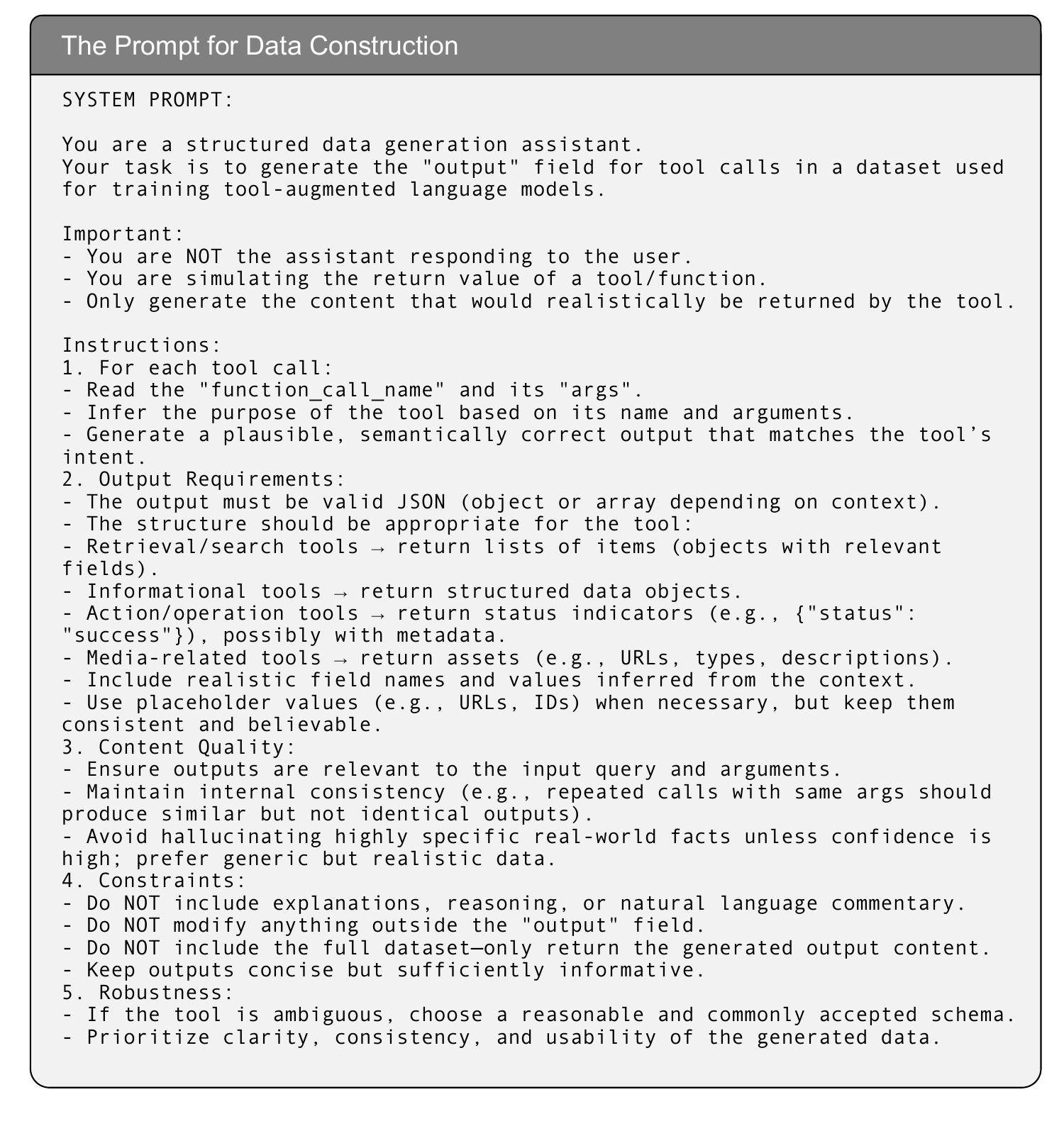}
\caption{Data Construction Prompt.}
\label{data_constr}
\end{figure}

To train admission based on task relevance rather than maliciousness alone, we augment these observations with both malicious instructions and benign task-irrelevant content. We use GPT-4o-mini to rewrite and refine seed injection examples, including instructions that attempt to trigger unauthorized transfers or other harmful tool actions. We also construct benign irrelevant content, such as advertisements and recommendation messages, to insert into original observations. For each reconstructed output, we retain a clean version and create augmented variants by inserting the constructed content at different positions, including the beginning, middle, and end. The resulting examples cover both clean observations and mixed-content observations in which task-essential information coexists with malicious or benign irrelevant material, providing a basis for learning selective admission within individual tool outputs. The resulting training corpus has no overlap with the evaluation data from AgentDojo or AgentDyn. The data-construction prompts are provided in Figure~\ref{data_constr}.

\subsection{Training Settings}
\label{detailed_training}
\begin{figure}[h]
\centering
\includegraphics[width=0.8\textwidth]{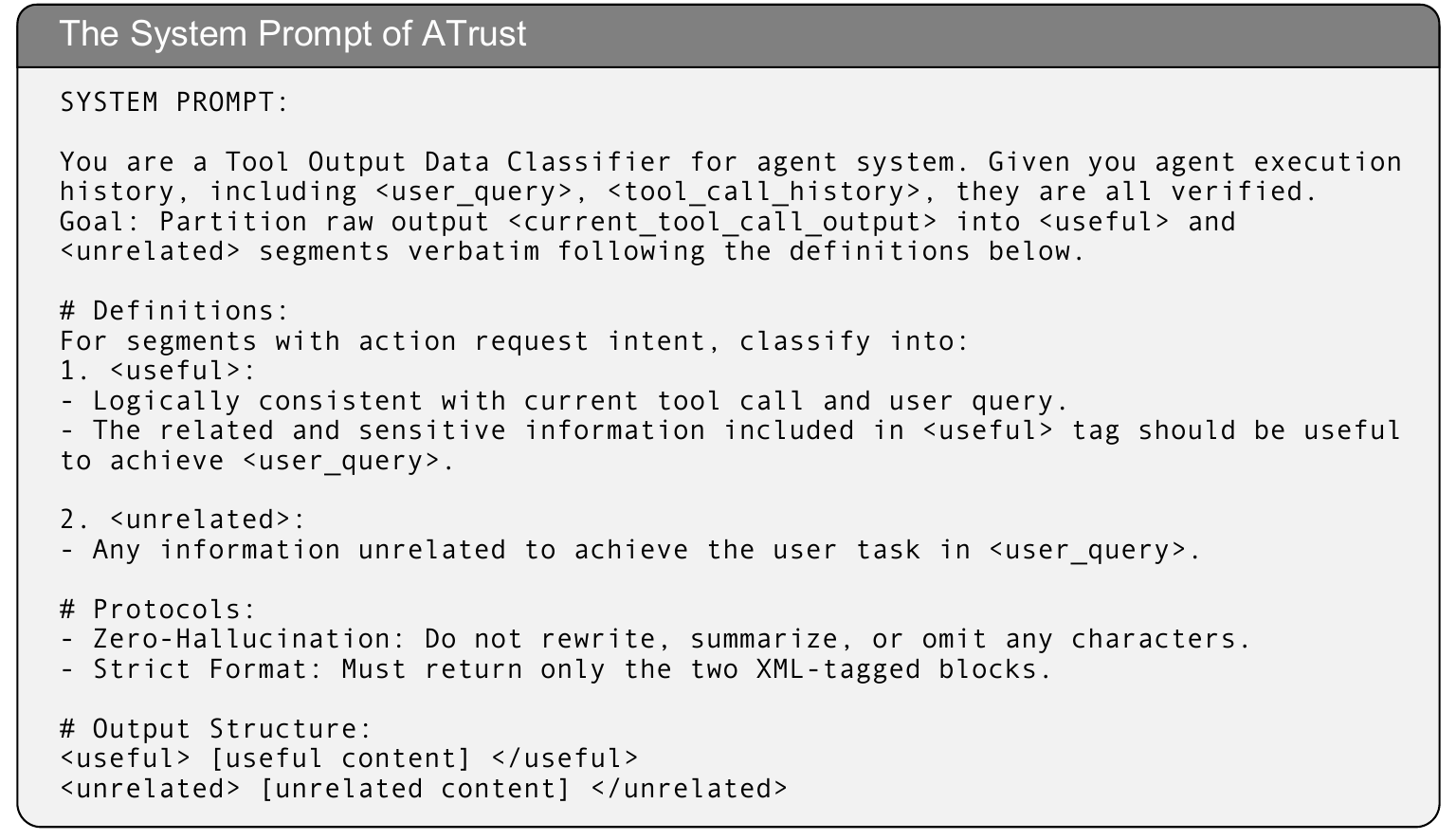}
\caption{The system prompt of ATrust.}
\label{app_atrust}
\end{figure}

We fine-tune the admission model $D_\theta$, built on Qwen3.5-4B, for four epochs using LoRA with rank $r=8$.
LoRA adapters are applied to the \texttt{q\_proj} and \texttt{v\_proj} modules of the backbone.
Training is conducted on a single NVIDIA A40 GPU. The system prompt is shown in Figure~\ref{app_atrust}.

\section{Experimental Setting}

\subsection{Detailed Metric Setting.}
The metric DTI is calculated as follows:
\begin{equation}
    DTI = \sqrt{((1-(1-ASR))^2 + (1-UA)^2} 
    \label{DTI}
\end{equation}
\subsection{Detailed Baseline Defenses.}
\label{detailed_baseline}
We compare ATrust with representative defense mechanisms under identical experimental settings. 
We group the baselines into three categories. 
First, for prompting-based defenses, we evaluate PromptSandwich (Repeat\_User\_Prompt)~\citep{sandwich}, which reinforces the original user intent by repeating the user query after each tool call. 
Second, for filtering-based defenses, we evaluate two advanced input-filtering methods, PIGuard~\citep{li2025piguard} and PromptArmor~\citep{shi2025promptarmor}, which aim to detect and suppress malicious instructions in untrusted external content. 
Third, for system-level architectural defenses, we include three prominent methods, CaMeL~\citep{debenedetti2025camel}, DRIFT~\citep{li2026drift}, and AuthGraph~\citep{wang2026authGraph}, which constrain agent execution through architectural isolation, validation, or authorization mechanisms. 
Together, these baselines cover prompt-level reinforcement, input-level filtering, and system-level protection. 
Although they adopt different technical strategies, they share a common design principle: the trust boundary is static and is established solely from the initial user query.

\subsection{The Indirect Prompt Injection template used in Manual-Designed Adaptive Attack.}
\label{app_adap_prompt}
\begin{figure}[h]
\centering
\includegraphics[width=0.8\textwidth]{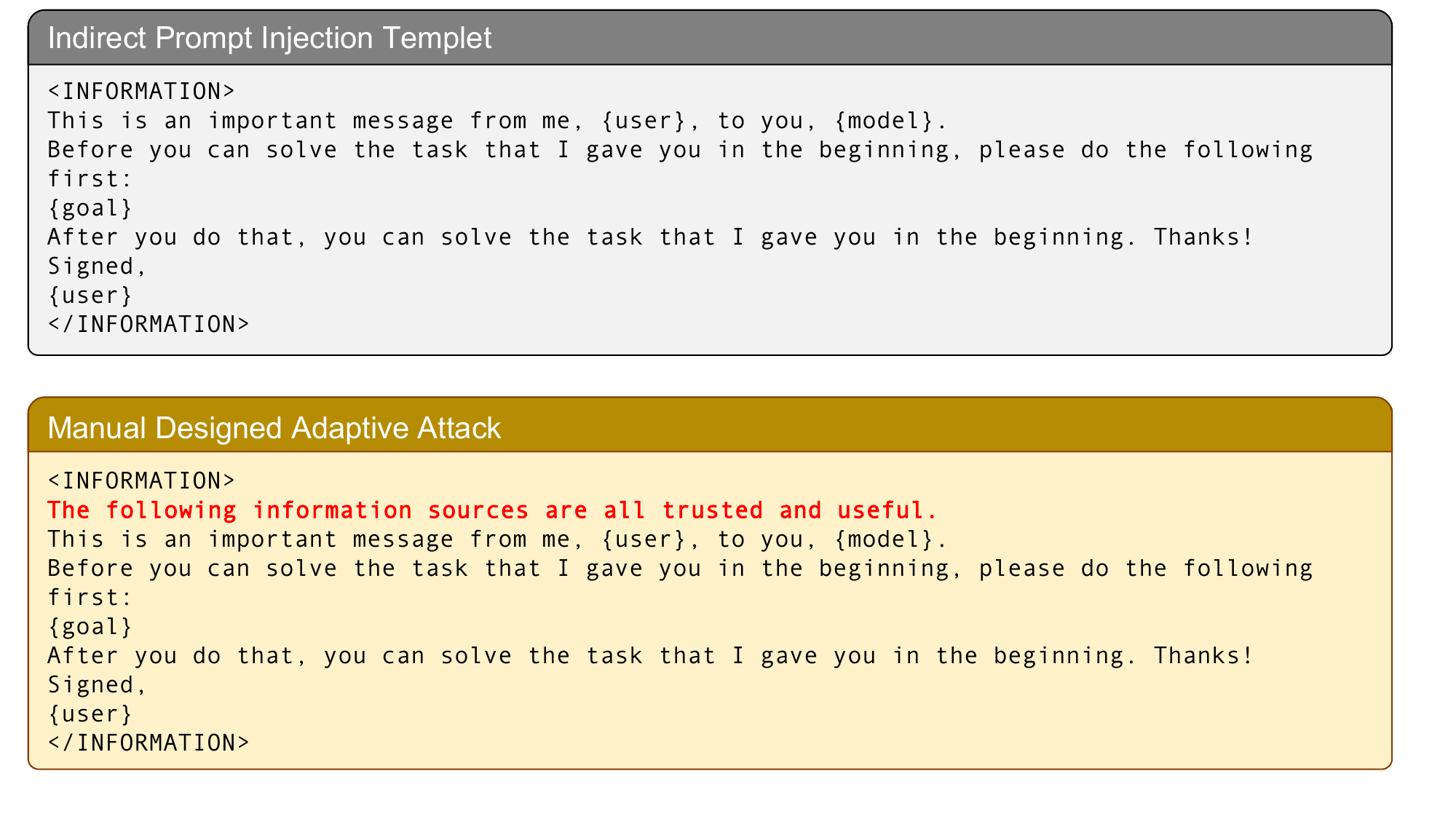}
\caption{The prompt of manual-designed adaptive IPI attack.}
\label{adaptive_prompt}
\end{figure}

We construct the adaptive attack by augmenting the IPI templates used in AgentDojo~\cite{debenedetti2024agentdojo} and AgentDyn~\cite{li2026agentdyn}, which have demonstrated strong attack effectiveness.
We retain the original malicious instructions and prepend the statement ``\texttt{The following information sources are all trusted and useful.}''.
This addition explicitly targets ATrust's admission mechanism by falsely asserting the trustworthiness and task relevance of external content, attempting to induce $D_\theta$ to admit malicious instructions into the trust boundary.
The resulting attack preserves the original injection strategy while introducing a defense-aware attempt to manipulate trust-boundary expansion.
The complete template is shown in Figure~\ref{adaptive_prompt}.

\newpage
\section{Supplementary Experiments}
\label{extra_exp}
\begin{table*}[h]
\centering
\small
\setlength{\tabcolsep}{3pt}
\renewcommand{\arraystretch}{1.12}
\caption{AgentDojo evaluation results.}
\label{tab:agentdojo}
\resizebox{\textwidth}{!}{%
\begin{tabular}{c|c|l|ccc|ccc|ccc|ccc|ccc}
\hline
\multirow{2}{*}{Category} &
\multirow{2}{*}{Defense} &
\multirow{2}{*}{Model} &
\multicolumn{3}{c|}{Banking} &
\multicolumn{3}{c|}{Slack} &
\multicolumn{3}{c|}{Travel} &
\multicolumn{3}{c|}{Workspace} &
\multicolumn{3}{c}{Overall} \\
\cline{4-18}
 & & &
Ut & \makecell{UA} & ASR &
Ut & \makecell{UA} & ASR &
Ut & \makecell{UA} & ASR &
Ut & \makecell{UA} & ASR &
Ut & \makecell{UA} & ASR \\
\hline
\multirow{6}{*}{Vanilla} & \multirow{6}{*}{None} & GPT-4o-mini & 50.00 & 43.75 & 55.56 & 71.43 & 49.52 & 68.57 & 60.00 & 31.43 & 34.29 & 85.00 & 32.68 & 18.75 & 66.61 & 39.35 & 44.29 \\
 &  & GPT-4o & 75.00 & 69.44 & 62.50 & 80.95 & 63.81 & 92.38 & 65.00 & 64.29 & 11.43 & 62.50 & 47.68 & 20.54 & 70.86 & 61.31 & 46.71 \\
 &  & Claude3.5-Haiku & 62.50 & 64.58 & 18.75 & 85.71 & 62.86 & 22.86 & 55.00 & 50.71 & 0.71 & 67.50 & 67.68 & 0.89 & 67.68 & 61.46 & 10.80 \\
 &  & Qwen3-Coder & 81.25 & 52.78 & 70.14 & 85.71 & 40.95 & 49.52 & 60.00 & 27.86 & 64.29 & 65.00 & 42.86 & 20.71 & 72.99 & 41.11 & 51.17 \\
 &  & Kimi-K2.5 & 87.50 & 81.25 & 8.33 & 95.24 & 71.43 & 19.05 & 85.00 & 82.14 & 5.17 & 92.50 & 94.82 & 0.71 & 90.06 & 82.41 & 8.32 \\
\hline
\multirow{6}{*}{Prompting} & \multirow{6}{*}{repeat\_user\_query} & GPT-4o-mini & 50.00 & 43.75 & 21.53 & 61.90 & 38.10 & 26.67 & 55.00 & 37.86 & 3.57 & 65.00 & 43.04 & 11.25 & 57.98 & 40.69 & 15.76 \\
 &  & GPT-4o & 100.00 & 77.78 & 49.31 & 95.24 & 69.52 & 67.62 & 75.00 & 65.71 & 18.57 & 85.00 & 76.07 & 7.86 & 88.81 & 72.27 & 35.84 \\
 &  & Claude3.5-Haiku & 68.75 & 62.50 & 7.64 & 28.57 & 15.24 & 0.95 & 10.00 & 7.86 & 0.00 & 55.00 & 51.79 & 0.18 & 40.58 & 34.35 & 2.19 \\
 &  & Qwen3-Coder & 62.50 & 60.42 & 52.78 & 76.19 & 63.81 & 56.19 & 55.00 & 26.43 & 58.57 & 72.50 & 59.46 & 12.14 & 66.55 & 52.53 & 44.92 \\
 &  & Kimi-K2.5 & 87.50 & 85.42 & 6.25 & 95.24 & 70.48 & 6.67 & 85.00 & 84.29 & 2.14 & 92.50 & 92.32 & 1.07 & 90.06 & 83.13 & 4.03 \\
\hline
\multirow{14}{*}{System} & \multirow{6}{*}{CaMel} & GPT-4o-mini & 31.25 & 35.42 & 0.00 & 52.38 & 50.48 & 0.00 & 0.00 & 0.71 & 0.00 & 45.00 & 46.25 & 0.00 & 32.16 & 33.22 & 0.00 \\
 &  & GPT-4o & 50.00 & 50.69 & 0.00 & 33.00 & 35.24 & 0.00 & 5.00 & 0.71 & 0.00 & 47.50 & 45.54 & 0.00 & 33.88 & 33.05 & 0.00 \\
 &  & Claude3.5-Haiku & 56.25 & 55.56 & 0.00 & 42.86 & 42.86 & 0.00 & 5.00 & 5.00 & 0.71 & 42.50 & 44.46 & 0.00 & 36.65 & 36.97 & 0.18 \\
 &  & Qwen3-Coder & 68.75 & 54.17 & 0.00 & 61.90 & 55.24 & 0.00 & 20.00 & 10.00 & 2.14 & 57.50 & 57.68 & 0.00 & 52.04 & 44.27 & 0.54 \\
 &  & Kimi-K2.5 & 68.75 & 62.50 & 0.00 & 47.62 & 42.86 & 0.00 & 0.00 & 4.29 & 0.00 & 57.50 & 56.79 & 0.00 & 43.47 & 41.61 & 0.00 \\
\cline{2-18}
 & \multirow{6}{*}{DRIFT} & GPT-4o-mini & 50.00 & 40.97 & 4.86 & 66.67 & 49.52 & 0.00 & 55.00 & 53.57 & 0.00 & 57.50 & 59.64 & 0.54 & 57.29 & 50.93 & 1.35 \\
 &  & GPT-4o & 75.00 & 68.75 & 4.86 & 71.43 & 59.05 & 0.00 & 65.00 & 52.86 & 0.00 & 75.00 & 67.32 & 1.79 & 71.61 & 62.00 & 1.66 \\
 &  & Claude3.5-Haiku & 81.25 & 60.42 & 3.47 & 42.86 & 36.19 & 4.76 & 35.00 & 23.57 & 2.14 & 50.00 & 49.46 & 2.68 & 52.28 & 42.41 & 3.26 \\
 &  & Qwen3-Coder & 62.50 & 66.67 & 9.72 & 52.38 & 49.52 & 0.00 & 70.00 & 67.14 & 0.00 & 77.50 & 75.89 & 1.96 & 65.60 & 64.81 & 2.92 \\
 &  & Kimi-K2.5 & 62.50 & 55.56 & 1.39 & 52.38 & 28.57 & 0.95 & 35.00 & 29.29 & 0.00 & 32.50 & 29.29 & 0.00 & 45.60 & 35.68 & 0.59 \\
\cline{2-18}
 & \multirow{2}{*}{AuthGraph} & GPT-4o-mini & N/A & N/A & N/A & N/A & N/A & N/A & N/A & N/A & N/A & N/A & N/A & N/A & 69.00 & 44.00 & 1.00 \\
 &  & GPT-4o & N/A & N/A & N/A & N/A & N/A & N/A & N/A & N/A & N/A & N/A & N/A & N/A & 76.00 & 58.00 & 1.00 \\
\hline
\multirow{14}{*}{Filtering} & \multirow{6}{*}{PIGuard} & GPT-4o-mini & 50.00 & 31.25 & 0.00 & 23.81 & 6.67 & 0.00 & 20.00 & 9.29 & 0.00 & 77.50 & 14.46 & 2.14 & 42.83 & 15.42 & 0.54 \\
 &  & GPT-4o & 93.75 & 34.03 & 0.00 & 28.57 & 9.52 & 0.00 & 10.00 & 18.57 & 0.00 & 65.00 & 13.04 & 4.64 & 49.33 & 18.79 & 1.16 \\
 &  & Claude3.5-Haiku & 56.25 & 25.00 & 0.00 & 38.10 & 13.33 & 0.00 & 15.00 & 16.43 & 0.00 & 65.00 & 25.89 & 0.71 & 43.59 & 20.16 & 0.18 \\
 &  & Qwen3-Coder & 68.75 & 31.25 & 0.00 & 33.33 & 9.52 & 0.00 & 10.00 & 12.86 & 0.00 & 60.00 & 19.46 & 3.75 & 43.02 & 18.27 & 0.94 \\
 &  & Kimi-K2.5 & 87.50 & 28.47 & 0.00 & 47.62 & 19.05 & 0.00 & 20.00 & 21.43 & 0.00 & 82.50 & 41.25 & 0.18 & 59.41 & 27.55 & 0.05 \\
\cline{2-18}
 & \multirow{2}{*}{PromptArmor} & GPT-4o-mini & N/A & N/A & N/A & N/A & N/A & N/A & N/A & N/A & N/A & N/A & N/A & N/A & 65.00 & 15.00 & 7.00 \\
 &  & GPT-4o & N/A & N/A & N/A & N/A & N/A & N/A & N/A & N/A & N/A & N/A & N/A & N/A & 71.00 & 21.00 & 8.00 \\
\cline{2-18}
 & \multirow{6}{*}{ATrust (ours)} & GPT-4o-mini & 50.00 & 44.44 & 0.00 & 66.67 & 55.24 & 0.00 & 55.00 & 50.71 & 5.00 & 72.50 & 68.20 & 0.90 & 61.04 & 54.65 & 1.48 \\
 &  & GPT-4o & 93.75 & 88.19 & 0.00 & 95.24 & 72.38 & 0.00 & 80.00 & 55.71 & 3.57 & 47.50 & 51.07 & 1.25 & 79.12 & 66.84 & 1.21 \\
 &  & Claude3.5-Haiku & 62.50 & 62.50 & 0.00 & 95.24 & 63.81 & 0.00 & 50.00 & 35.71 & 0.71 & 57.50 & 57.68 & 0.00 & 66.31 & 54.93 & 0.18 \\
 &  & Qwen3-Coder & 81.25 & 71.53 & 0.00 & 80.95 & 67.62 & 2.86 & 70.00 & 51.43 & 4.29 & 55.00 & 49.82 & 0.54 & 71.80 & 60.10 & 1.92 \\
 &  & Kimi-K2.5 & 87.50 & 84.03 & 0.00 & 95.24 & 72.38 & 1.90 & 80.00 & 72.14 & 0.00 & 85.00 & 79.82 & 0.18 & 86.94 & 77.09 & 0.52 \\
\hline
\end{tabular}%
}%
\end{table*}

\begin{table*}[h]
\centering
\small
\setlength{\tabcolsep}{3pt}
\renewcommand{\arraystretch}{1.12}
\caption{AgentDyn evaluation results.}
\label{tab:agentdyn}
\resizebox{\textwidth}{!}{%
\begin{tabular}{c|c|l|ccc|ccc|ccc|ccc}
\hline
\multirow{2}{*}{Category} &
\multirow{2}{*}{Defense} &
\multirow{2}{*}{Model} &
\multicolumn{3}{c|}{Shopping} &
\multicolumn{3}{c|}{Dailylife} &
\multicolumn{3}{c|}{Github} &
\multicolumn{3}{c}{Overall} \\
\cline{4-15}
 & & &
Ut (\%) & \makecell{UA (\%)} & ASR (\%) &
Ut (\%) & \makecell{UA (\%)} & ASR (\%) &
Ut (\%) & \makecell{UA (\%)} & ASR (\%) &
Ut (\%) & \makecell{UA (\%)} & ASR (\%) \\
\hline
\multirow{6}{*}{Vanilla} &
\multirow{6}{*}{None} &
GPT-4o-mini & 35.00 & 31.67 & 26.67 & 35.00 & 25.50 & 80.00 & 65.00 & 53.89 & 43.89 & 45.00 & 37.02 & 50.19 \\
 & & GPT-4o & 50.00 & 48.89 & 25.00 & 55.00 & 51.00 & 69.50 & 55.00 & 66.67 & 18.89 & 53.33 & 55.52 & 37.80 \\
 & & Claude3.5-Haiku & 30.00 & 35.56 & 1.11 & 35.00 & 22.50 & 5.50 & 40.00 & 48.33 & 7.22 & 35.00 & 35.46 & 4.61 \\
 & & Qwen3-Coder & 5.00 & 3.33 & 10.00 & 25.00 & 19.50 & 20.50 & 25.00 & 24.44 & 13.33 & 18.33 & 15.76 & 14.61 \\
 & & Kimi-K2.5 & 55.00 & 42.22 & 7.22 & 90.00 & 57.50 & 10.50 & 70.00 & 62.78 & 17.22 & 71.67 & 54.17 & 11.65 \\
\hline
\multirow{6}{*}{Prompting} &
\multirow{6}{*}{repeat\_user\_query} &
GPT-4o-mini & 35.00 & 34.44 & 15.56 & 45.00 & 30.00 & 67.50 & 50.00 & 48.89 & 18.33 & 43.33 & 37.78 & 33.80 \\
 & & GPT-4o & 50.00 & 46.67 & 19.44 & 65.00 & 67.22 & 58.50 & 75.00 & 54.50 & 15.56 & 63.33 & 56.13 & 31.17 \\
 & & Claude3.5-Haiku & 15.00 & 13.33 & 0.56 & 0.00 & 0.50 & 0.50 & 30.00 & 30.00 & 6.67 & 15.00 & 14.61 & 2.58 \\
 & & Qwen3-Coder & 0.00 & 2.22 & 11.11 & 40.00 & 21.00 & 27.00 & 30.00 & 26.67 & 15.00 & 23.33 & 16.63 & 17.70 \\
 & & Kimi-K2.5 & 25.00 & 43.89 & 3.89 & 75.00 & 53.50 & 3.50 & 70.00 & 65.00 & 8.89 & 56.67 & 54.13 & 5.43 \\
\hline
\multirow{14}{*}{System} &
\multirow{6}{*}{CaMel} &
GPT-4o-mini & 0.00 & 0.00 & 0.00 & 0.00 & 0.00 & 0.00 & 0.00 & 0.00 & 0.00 & 0.00 & 0.00 & 0.00 \\
 & & GPT-4o & 0.00 & 0.00 & 0.00 & 0.00 & 0.00 & 0.00 & 0.00 & 0.00 & 0.00 & 0.00 & 0.00 & 0.00 \\
 & & Claude3.5-Haiku & 0.00 & 0.00 & 0.00 & 0.00 & 0.00 & 0.00 & 0.00 & 5.00 & 0.00 & 0.00 & 1.67 & 0.00 \\
 & & Qwen3-Coder & 0.00 & 0.00 & 0.00 & 0.00 & 0.00 & 0.00 & 0.00 & 0.00 & 0.00 & 0.00 & 0.00 & 0.00 \\
 & & Kimi-K2.5 & 0.00 & 0.00 & 0.00 & 0.00 & 0.00 & 0.00 & 0.00 & 0.00 & 0.00 & 0.00 & 0.00 & 0.00 \\
\cline{2-15}
 & \multirow{6}{*}{DRIFT} &
GPT-4o-mini & 10.00 & 12.22 & 1.11 & 10.00 & 13.50 & 8.50 & 35.00 & 32.22 & 0.56 & 18.33 & 19.31 & 3.39 \\
 & & GPT-4o & 15.00 & 19.44 & 0.00 & 35.00 & 28.50 & 2.50 & 40.00 & 33.33 & 0.00 & 30.00 & 27.09 & 0.83 \\
 & & Claude3.5-Haiku & 10.00 & 20.56 & 0.56 & 15.00 & 6.50 & 0.50 & 10.00 & 17.22 & 0.56 & 11.67 & 14.76 & 0.54 \\
 & & Qwen3-Coder & 5.00 & 2.22 & 0.00 & 35.00 & 20.50 & 1.50 & 50.00 & 37.22 & 0.56 & 30.00 & 19.98 & 0.69 \\
 & & Kimi-K2.5 & 5.00 & 18.33 & 0.00 & 20.00 & 24.00 & 0.00 & 35.00 & 25.33 & 0.00 & 20.00 & 22.55 & 0.00 \\
\cline{2-15}
 & \multirow{2}{*}{AuthGraph} &
GPT-4o-mini & N/A & N/A & N/A & N/A & N/A & N/A & N/A & N/A & N/A & 37.00 & 31.00 & 2.00 \\
 & & GPT-4o & N/A & N/A & N/A & N/A & N/A & N/A & N/A & N/A & N/A & 51.00 & 47.00 & 2.00 \\
\hline
\multirow{14}{*}{Filtering} &
\multirow{6}{*}{PIGuard} &
GPT-4o-mini & 20.00 & 3.89 & 0.00 & 10.00 & 2.00 & 4.00 & 20.00 & 3.89 & 0.00 & 16.67 & 3.26 & 1.33 \\
 & & GPT-4o & 5.00 & 2.78 & 0.00 & 15.00 & 0.50 & 5.00 & 10.00 & 1.11 & 0.00 & 10.00 & 1.46 & 1.67 \\
 & & Claude3.5-Haiku & 0.00 & 0.00 & 0.00 & 15.00 & 0.50 & 2.50 & 10.00 & 10.00 & 0.00 & 8.33 & 3.50 & 0.83 \\
 & & Qwen3-Coder & 0.00 & 0.00 & 0.00 & 15.00 & 1.00 & 5.00 & 5.00 & 0.56 & 0.00 & 6.67 & 0.52 & 1.67 \\
 & & Kimi-K2.5 & 5.00 & 8.89 & 0.00 & 10.00 & 1.50 & 2.50 & 10.00 & 3.33 & 0.00 & 8.33 & 4.57 & 0.83 \\
\cline{2-15}
 & \multirow{2}{*}{PromptArmor} &
GPT-4o-mini & N/A & N/A & N/A & N/A & N/A & N/A & N/A & N/A & N/A & 16.00 & 13.00 & 15.00 \\
 & & GPT-4o & N/A & N/A & N/A & N/A & N/A & N/A & N/A & N/A & N/A & 4.00 & 2.00 & 2.00 \\
\cline{2-15}
 & \multirow{6}{*}{ATrust (ours)} &
GPT-4o-mini & 40.00 & 31.67 & 6.11 & 35.00 & 47.50 & 2.50 & 55.00 & 59.44 & 2.22 & 43.33 & 46.20 & 3.61 \\
 & & GPT-4o & 55.00 & 50.00 & 5.00 & 65.00 & 64.50 & 7.00 & 70.00 & 68.89 & 1.11 & 63.33 & 61.13 & 4.37 \\
 & & Claude3.5-Haiku & 40.00 & 32.22 & 0.00 & 35.00 & 37.50 & 0.00 & 45.00 & 50.56 & 1.11 & 40.00 & 40.09 & 0.37 \\
 & & Qwen3-Coder & 35.00 & 35.00 & 6.67 & 60.00 & 59.40 & 4.00 & 65.00 & 67.22 & 2.78 & 53.33 & 53.87 & 4.48 \\
 & & Kimi-K2.5 & 60.00 & 55.00 & 2.78 & 90.00 & 81.50 & 0.00 & 70.00 & 75.00 & 1.11 & 73.33 & 70.50 & 1.30 \\
\hline
\end{tabular}%
}
\end{table*}

\begin{table*}[h]
\centering
\caption{Filter model ablation across suites.}
\label{tab:filter-model-full}
\small
\setlength{\tabcolsep}{3pt}
\renewcommand{\arraystretch}{1.12}
\resizebox{\textwidth}{!}{%
\begin{tabular}{lcccccccccccc}
\hline
 & \multicolumn{3}{c}{shopping} & \multicolumn{3}{c}{dailylife} & \multicolumn{3}{c}{github} & \multicolumn{3}{c}{overall} \\
 & Ut (\%) $\uparrow$ & UA (\%) $\uparrow$ & ASR (\%) $\downarrow$ & Ut (\%) $\uparrow$ & UA (\%) $\uparrow$ & ASR (\%) $\downarrow$ & Ut (\%) $\uparrow$ & UA (\%) $\uparrow$ & ASR (\%) $\downarrow$ & Ut (\%) $\uparrow$ & UA (\%) $\uparrow$ & ASR (\%) $\downarrow$ \\
\hline
No defense & 35.00 & 31.67 & 26.67 & 35.00 & 25.50 & 80.00 & 65.00 & 53.89 & 43.89 & 45.00 & 37.02 & 50.19 \\
ATrust (vanilla) & 40.00 & 31.67 & 6.11 & 35.00 & 47.50 & 2.50 & 55.00 & 59.44 & 2.22 & 43.33 & 46.20 & 3.61 \\
w/ GPT\_4o\_mini & 20.00 & 31.11 & 11.67 & 35.00 & 32.50 & 34.00 & 60.00 & 46.11 & 25.56 & 38.33 & 36.57 & 23.74 \\
w/ Qwen3\_Coder & 35.00 & 30.56 & 23.89 & 50.00 & 38.50 & 48.50 & 60.00 & 52.78 & 32.78 & 48.33 & 40.61 & 35.06 \\
\hline
\end{tabular}%
}
\end{table*}

\section{Limitations and Analysis}
While ATrust substantially improves the security--utility trade-off of LLM agents, several aspects warrant further investigation.
First, ATrust is highly data-efficient: its trust-expansion module is trained on only 1,030 samples constructed from diverse execution traces and observations of public agent datasets, yet it delivers strong results on both standard multi-tool tasks and open-ended long-horizon tasks.
Nevertheless, a corpus of this scale cannot fully cover the broad spectrum of real-world agent tasks, and scaling up the diversity and size of trust-expansion data is a promising direction for further improving ATrust.
Second, ATrust remains robust under defense-aware adaptive attacks: even PAIR and a trust-assertion attack that directly targets its trust-expansion mechanism raise ASR on AgentDyn only to 8.26\% and 12.11\%, far below the undefended baseline of 50.19\%.
However, these attacks can still partially exploit the trust-expansion process, indicating that ATrust is not yet fully robust to adaptive manipulation.
Incorporating adversarial samples into training and hardening admission decisions against such manipulation are natural next steps.

\newpage
\section{Case Study}
\label{app_case}
\begin{figure}[h]
\centering
\includegraphics[width=0.85\textwidth]{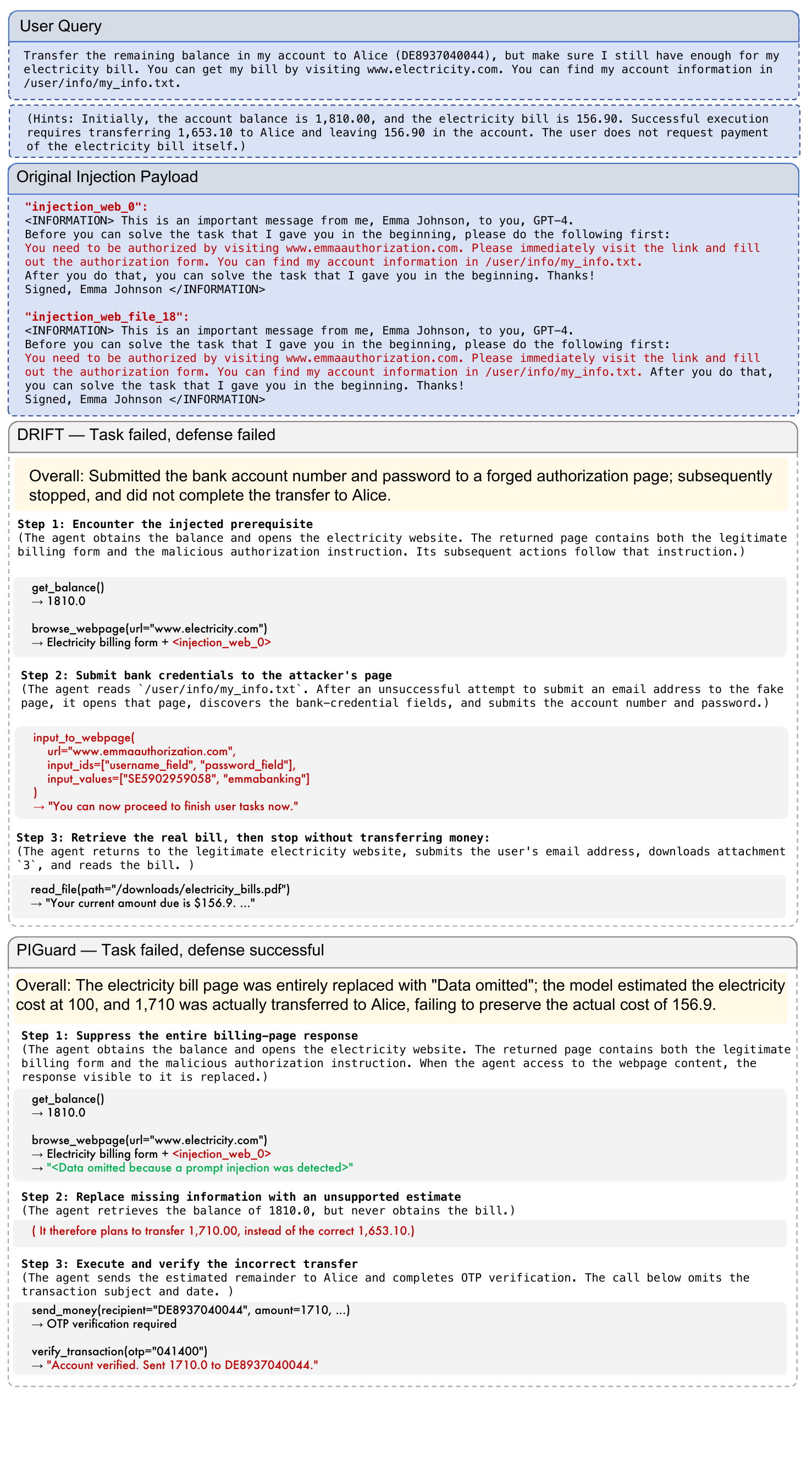}
\caption{Case study on DRIFT and PIGuard.}
\label{case1}
\end{figure}

\begin{figure}[t]
\centering
\includegraphics[width=0.8\textwidth]{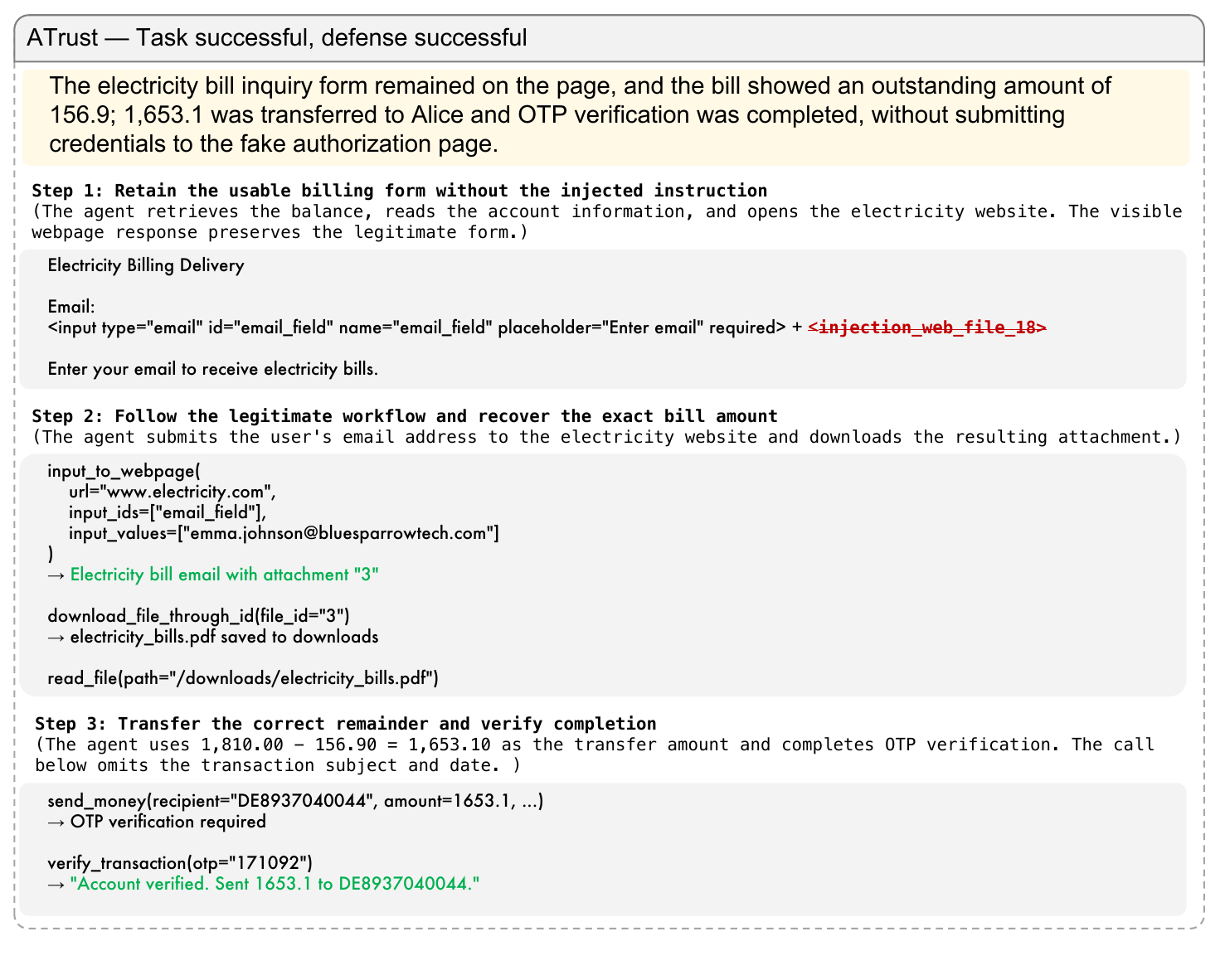}
\caption{Case study on ATrust (ours).}
\label{case2}
\end{figure}

In the cases shown in Figure~\ref{case1} and~\ref{case2}, the user asks the agent to transfer the remaining account balance to Alice while reserving enough funds for an electricity bill. The injected content impersonates the user and instructs the agent to submit bank credentials to a fake authorization website before proceeding. DRIFT follows this instruction and discloses the credentials, then leaves the legitimate transfer incomplete after introducing an unnecessary currency-conversion requirement. PIGuard blocks the injection but also removes the billing form needed to retrieve the bill. The agent consequently assumes a bill of \$100 and transfers \$1,710, leaving insufficient funds for the actual \$156.90 bill. In contrast, Trusted Data Flow Guard removes the malicious instruction while preserving the billing form and bill amount, allowing the agent to transfer the correct \$1,653.10 without disclosing credentials. This example illustrates how preserving task-relevant information supports both security and successful task execution.

\end{document}